\documentclass[aps,a4paper,10pt,onecolumn]{revtex4}
\usepackage[pdftex]{graphicx}	
	\DeclareGraphicsExtensions{.pdf,.png,.jpg}
\usepackage[applemac]{inputenc}
\usepackage{amsmath}
\usepackage{amssymb,amsthm}
\usepackage{mathbbol,bbm}
\usepackage{graphicx,float}
\usepackage[final]{showkeys}
\usepackage{bm,dsfont}
\usepackage{psfrag}

\def\ket#1{\left|#1\right>}
\def\bra#1{\left<#1\right|}
\def\Tr{ {\rm{Tr }}}

\begin{document}
\title{Symmetry-Protected Topological Phases at Finite Temperature}
\author{O. Viyuela, A. Rivas and M.A. Martin-Delgado}
\affiliation{Departamento de F\'{\i}sica Te\'orica I, Universidad Complutense, 28040 Madrid, Spain}

\vspace{-3.5cm}

\begin{abstract}
We have applied the recently developed theory of topological Uhlmann numbers to a representative model of a
topological insulator in two dimensions, the Qi-Wu-Zhang model. We have found a stable symmetry-protected topological (SPT) phase under external thermal fluctuations in two-dimensions. A complete phase diagram for this model is computed as a function of
temperature and coupling constants in the original Hamiltonian. It shows the appearance of large stable phases of matter
with topological properties compatible with thermal fluctuations or external noise and the existence of critical lines separating
abruptly trivial phases from topological phases. These novel critical temperatures represent thermal topological phase transitions.
The initial part of the paper comprises a self-contained explanation of the Uhlmann geometric phase needed to understand
the topological properties that it may acquire when applied to topological insulators and superconductors.
\end{abstract}



\maketitle

\tableofcontents

\section{INTRODUCTION}
\label{sec:I}

Topological insulators and superconductors \cite{Haldane_88,kane_mele05,bernevig_zhang06,kane_mele05b,moore_balents07,fu_kane_mele07,rmp1,rmp2,LibroBernevig} are novel phases of matter that cannot be described using the standard Gizburg-Landau theory for phase transitions. These novel phases are of great interest for applications in emerging fields such as spintronics, photonics or quantum computing. They have opened an area of new rich physical phenomenology, giving birth to materials that host robust metallic edge states and the possibility for
realising  Majorana fermions \cite{Kitaev01,Volovik_99,Read_00,Ivanov_01}, as well as a zoo of new exotic particles known as ‘anyons’ \cite{LM77, Wilczek82, ASW85, Nayak_08}. Actually, graphene was the first material proposed to realise a topological insulating phase using a mechanism of spin-orbital coupling between two graphene sheets \cite{kane_mele05}. Unfortunately, the gap due to this effect in carbon atoms is too small and the quantum spin Hall effect was finally achieved in HgTe quantum wells instead \cite{Bernevig_06,Konig_07}. Very recently,  bosonic topological orders have been also proposed \cite{VS13,MKF13,Senthil14,YW13,wen_etal14}.

Whereas these systems have been extensively studied for the idealized case of zero temperature, the common belief was that these exotic phases are gradually spoiled as the systems are heated up \cite{Viyuela_et_al12,Mazza_et_al12}. In \cite{Viyuela_et_al12}, it was shown that certain topological properties like the edge states can be destroyed by a thermal bath even when the system-bath interaction respect the protecting symmetry of the topological insulator. It was not until the work in \cite{Rivas_et_al13} that a topological indicator with a dependence on the temperature of the system was introduced, that was related to the topological part of the quantum Hall conductivity and the degree of mixedness of the edge states due to the dissipative effect. Other attempts to understand the interplay between topological phases and finite temperature \cite{Garate13,Garate14}, nonequilibrium dynamics \cite{Evert} and dissipation-engeneering \cite{Bardyn_et_al12,Kraus_et_al12,Bardyn_et_al13,Budich_et_al_14} have also been made.

The extension of the notion of symmetry-protected topological (SPT) order for fermions to situations where the system is not in a pure state,
but in a quantum mixed state described by a density matrix has been developed very recently. The first realization of topological
fermonic phases in the presence of thermal effects or external noise was proposed in Ref. \cite{Viyuela_et_al14}.
Here a topological invariant remains quantized for a finite range of temperature/noise for the first time in emblematic examples
of topological insulators and superconductors in 1D. Then, new advances have been achieved for fermionic phases of matter
with SPT order in two-dimensional systems  \cite{Arovas14,Viyuela_et_al14_2D}.
 The novelty of these works relies on the introduction of two new observables: the Uhlmann phase for wire-shaped systems (1D) \cite{Viyuela_et_al14}, and the Uhlmann number for planar ones (2D) \cite{Viyuela_et_al14_2D} as well as a variety of extensions  \cite{Arovas14}, which are able to account for topological order even at non-zero temperature.
 Remarkably, these recent works report purely ``thermal" topological phase transitions between distinct non-trivial topological regions solely driven by temperature, without losing their quantum character. It is also shown that there exists always a critical temperature above which the topological phase abruptly disappears. The possibility of having several topological thermal critical points
has also been discovered as a novel effect \cite{Viyuela_et_al14_2D}.

In this way, exotic topological quantum properties can survive even in the presence of dissipative effects and finite temperature, which are the natural conditions in real life experiments. This situation for fermion systems with SPT order
is in sharp contrast to some results establishing the absence of stable topological phases subject to thermal effects, like for topological orders with spins \cite{AFH,Nussinov_09,Hastings,Selfcorrecting,Top_qutrits}, for SPT orders with spins \cite{Evert},
or SPT orders with fermions under certain conditions \cite{Viyuela_et_al12}.
Therefore, these new results open the way towards the characterization of  fermion quantum phases of matter with topological phases
protected by symmetry in thermal states, or more general density matrices. Note that other studies on thermal effects and related issues in topological systems have been recently carried out \cite{Albert_et_al14,Khoury_et_al14,Dauphin_et_al14,Shen_et_al14,Dehghani_et_al14,Sirker_et_al14,Ho_et_al14,Patrascu14,Yang_et_al14,Budich_et_al15}.

The aim of this paper is to give a complete and comprehensive derivation of the Uhlmann geometric phase as a tool to characterise symmetry protected topological orders at finite temperature, performing a detailed analysis of its intrinsic properties. Firstly, we will formally derive the expression for the Uhlmann phase as an extension of the Berry phase to the case of dealing with quantum
mixed states described by density matrices. Later, we apply the formalism to new 2D models for topological insulators with time-reversal broken symmetry (TRB) that can be characterized using these Uhlmann geometric concepts by the so-called topological Uhlmann numbers \cite{Viyuela_et_al14_2D}. 

The paper is organized as follows. Sec.~\ref{sec:II} introduces the concept of Uhlmann holonomies, giving a detailed derivation of the Uhlmann parallel transport condition and the geometric Uhlmann phase, by means of amplitudes and purifications of a density matrix. In Sec.~\ref{sec:III}, the topological Uhlmann phase in 1D is presented and applied to a generic 1D two-band model for a topological insulator or superconductor. In Sec.~\ref{sec:IV}, we derive the topological Uhlmann number for 2D topological systems, discussing some technical issues and using it to characterise an emblematic 2D topological insulator, namely the Qi-Wu-Zhang model \cite{QWZ_06,QHZ_08}. Sec.~\ref{sec:V} is devoted to conclusions.

\section{UHLMANN HOLONOMIES}
\label{sec:II}

Geometric phases have played an essential role in many quantum phenomena since its modern discovery by Berry \cite{Berry84} (see also Refs. \cite{Simon85,WilczekBook}). An emblematic example is the description of the quantum Hall effect using the theory of fiber bundles. More concretely, one can relate the transversal conductivity $\sigma_{xy}$ of the system with the first Chern number ${\rm Ch}$ as \cite{TKNN,Hatsugai},
\begin{equation}
\sigma_{xy}={\rm Ch}\frac{e^2}{h},
\label{conductivity}
\end{equation}
where $e$ is the electron  charge and $h$ is the Planck constant.

The Chern numbers are topological invariants associated to characteristic classes of complex vector bundles when working on an oriented manifold of even dimension. The first Chern number ${\rm Ch}$, which is defined for the 2D case, is just the integral of the Berry curvature over the two-dimensional parameter space, i.e. the Brillouin zone (BZ) in the context of condensed matter systems. In addition, this topological observable can also be related to the presence of conducting edge states when the system has open boundary conditions.

The standard way to observe the properties and implications that the Berry phase has in solids (Zak phase \cite{Zak_89}), has been through indirect measurements like the transverse conductivity, or side effects like the presence of conducting edge states at the boundary. However, very recently the first direct experimental measurement of a Berry phase in a cold atoms system has been achieved  \cite{Atala_et_al_12,Abanin_et_al_13,Demler_etal14}. In particular, \cite{Atala_et_al_12} reports the measurement of the Zak phase in a one-dimensional optical lattice simulating the different phases of polyacetylene, characterizing the topology of the system and observing the quantization of this phase. This paves the way towards the extension of geometric phases to study topological properties beyond the realm of condensed-matter systems \cite{Tarruell_et_al12,Jotzu_14, Duca_15,Aidelsburger_15}. Moreover, there are recent experiments on this context involving architectures of superconducting qubits \cite{Schroer_et_al14,Roushan_et_al14}.

Although these studies provide a successful picture for quantum systems in pure states, typically the ground state, very little is known about the fate of those topological phases of matter when the system is in a mixed quantum state represented by a density matrix.

A fundamental problem in the theory and applications of geometrical phases is the extension from pure quantum states (Berry) to mixed quantum states described by density matrices. Uhlmann was first to mathematically address this issue \cite{Uhlmann} and to provide a satisfactory solution \cite{Uhlmann2, Uhlmann3, Hubner, Uhlmann6}. For more than a decade, there has been a renewed interest in studying geometric phases for mixed states and under dissipative evolutions from the point of view of quantum information \cite{Soqvist_et_al_00,Anandan_et_al02,medidaU,Du_et_al03,Sjoqvist_07,Zhu_et_al11}. In the rest of the section, we will give a detailed derivation of the Uhlmann geometric phase and the definition of Uhlmann parallel transport. These two key ideas will allow us to construct later two topological observables to distinguish different topological phases in condensed matter systems both in 1D and 2D.

\subsection{Amplitudes and Purifications}

One of the most fundamental problems when applying geometric concepts to the set $\mathcal{Q}$ of density matrices is the nonlinear character of it. That is, consider some scalars $\lambda_1$ and $\lambda_2$, for all pairs of density matrices $\rho_1$ and $\rho_2$, the linear combination $\lambda_1\rho_1+\lambda_2\rho_2$ is not another density matrix unless $\lambda_1,\lambda_2\geq0$ and $\lambda_1+\lambda_2=1$. In the case of pure states $\rho=|\psi\rangle\langle\psi|$, this is not a major issue as we can immediately introduce a Hilbert space formed by the vectors $|\psi\rangle$ so that the quantum state $\rho=|\psi\rangle\langle\psi|$ is the projector on the subspace spanned by $|\psi\rangle$. Then the geometric properties of the Hilbert space can be applied to the vectors $|\psi\rangle$, but having in mind that $|\psi\rangle$ and ${\rm e}^{{\rm i} \phi}|\psi\rangle$ describe the same physical state $\rho=|\psi\rangle\langle\psi|$. Namely, to work with a Hilbert space we have to pay the price of introducing an additional degree of freedom, the phase of the vectors, such that quantum pure states as seen in the Hilbert space are defined modulo a ${\rm U}(1)$ phase factor.

The key idea is to realize that an analogous construction was possible for any general density matrix. Specifically, consider a density matrix $\rho$ and define its amplitude as any of the matrices $w$ such that
\begin{equation}\label{ww^daga}
\rho=ww^\dagger.
\end{equation}
The amplitudes form a Hilbert space $\mathcal{H}_w$ with the Hilbert-Schmidt product $(w_1,w_2)=\Tr(w_1^\dagger w_2)$. From Eq. \eqref{ww^daga}, we see that there is a ${\rm U}(n)$-gauge freedom in the choice of the amplitude (here $n$ denotes the dimension of the space), since $w$ and $wU$, where $U$ is a unitary operator, are amplitudes of the same state.

\begin{figure}[t]
\centering
\includegraphics[width=0.8\textwidth]{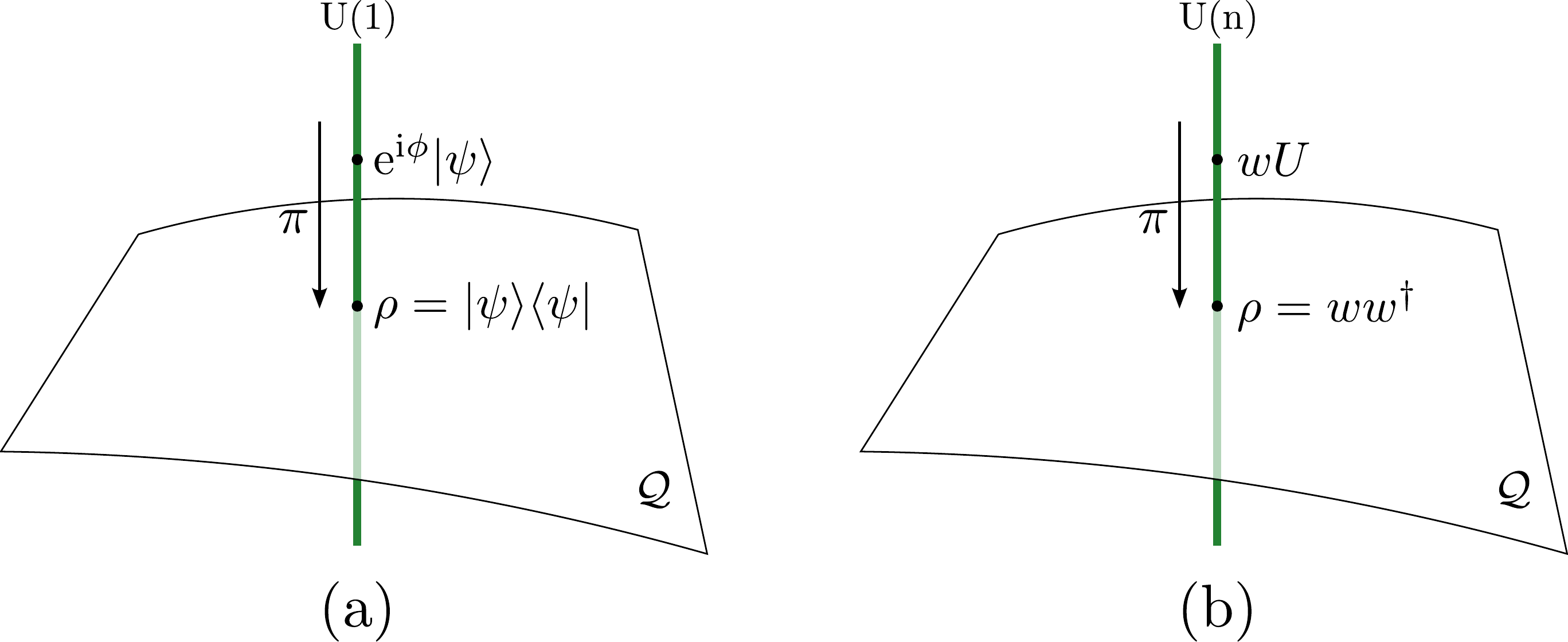}
\caption{Comparison between the usual fiber bundle of pure states (a), and the fiber bundle proposed by Uhlmann (b) for any density matrix through the concept of amplitude.}
\label{plotFibradoPuroMixto}
\end{figure}

Note the parallelism between the usual ${\rm U}(1)$-gauge freedom for pure states, and the ${\rm U}(n)$-gauge freedom for the amplitudes, see Fig. \ref{plotFibradoPuroMixto}. In the case of pure states, one defines a projection map $\pi:\mathcal{H}\mapsto\mathcal{Q}$:
\begin{equation}
\pi(|\psi\rangle)=|\psi\rangle\langle\psi|,
\end{equation}
so that $|\psi\rangle$ and ${\rm e}^{{\rm i}\phi}|\psi\rangle$ represent the same physical state, i.e. the same density matrix given by the projection map $\pi$. In the case of amplitudes, the projection map is analogously defined, $\pi:\mathcal{H}_w\mapsto\mathcal{Q}$:
\begin{equation}
\pi(w)=ww^\dagger.
\end{equation}
Thus, the usual gauge freedom for pure states can be seen as a particular case of the amplitude ${\rm U}(n)$-gauge freedom for general states.

So far the concept of amplitude seems just a mathematical abstraction, however it also enjoys a nice physical meaning. To see it, consider what is called the ``square root'' section of the Uhlmann fiber bundle, which is nothing but the parametrization $w=\sqrt{\rho}U$ for amplitudes of some density matrix $\rho$ (note that this is also called polar decomposition of $w$, see below). Then by using the spectral decomposition, $\rho=\sum_jp_j|\psi_j\rangle\langle\psi_j|$ we have that
\begin{equation}
w=\sum_j\sqrt{p_j}|\psi_j\rangle\langle\psi_j|U.
\end{equation}
Now we define the following isomorphism between the spaces $\mathcal{H}_w$ and $\mathcal{H}\otimes\mathcal{H}$,
\begin{equation}\label{isomorfismo}
w=\sum_j\sqrt{p_j}|\psi_j\rangle\langle\psi_j|U\longleftrightarrow |w\rangle=\sum_j\sqrt{p_j}|\psi_j\rangle\otimes U^{\rm t}|\psi_j\rangle,
\end{equation}
where $U^{\rm t}$ denotes the transposition of $U$ taken with respect to the eigenbasis of $\rho$. Then, the property $\rho=ww^\dagger$ is now written as
\begin{equation}\label{purification}
\rho=\Tr_2(|w\rangle\langle w|).
\end{equation}
Here, $\Tr_2$ denotes the partial trace over the second Hilbert space of $\mathcal{H}\otimes\mathcal{H}$. In other words, any amplitude $w$ of some density matrix $\rho$ can be seen as a pure state $|w\rangle$ of the enlarged space $\mathcal{H}\otimes\mathcal{H}$ with partial trace equals to $\rho$, Eq. \eqref{purification}. Thus, it is said that $|w\rangle$ is a purification of $\rho$. Thanks to this isomorphism between amplitudes and purifications the whole geometric approach by Uhlmann can be written in terms of purifications which are objects with a clear physical meaning. However, we shall keep on developing the theory and explanations using amplitudes and refer to the purification interpretation when needed.

\subsection{Parallel Amplitudes}

The concept of parallel amplitude is essential to define the Uhlmann parallel transport condition and the holonomy. Given a pair of states $\rho_1$ and $\rho_2$, we define two of their amplitudes $w_1$ and $w_2$ to be parallel if they minimize the Hilbert space distance in $\mathcal{H}_w$, i.e. $w_1\parallel w_2$ if
\begin{equation}
\| w_1-w_2 \|^2=\min_{\tilde{w}_1,\tilde{w}_2}\| \tilde{w}_1-\tilde{w}_2 \|^2, \quad \text{with } \rho_1=\tilde{w}_1\tilde{w}_1^\dagger, \ \rho_2=\tilde{w}_2\tilde{w}^\dagger_2.
\end{equation}
We may work further on this condition and establish some properties of such $w_1$ and $w_2$.
\begin{align}
\min_{\tilde{w}_1,\tilde{w}_2}\| \tilde{w}_1-\tilde{w}_2 \|^2&=\min_{\tilde{w}_1,\tilde{w}_2}\left(\tilde{w}_1-\tilde{w}_2,\tilde{w}_1-\tilde{w}_2\right)\nonumber \\
&=\min_{\tilde{w}_1,\tilde{w}_2}\Tr\left(\tilde{w}_1^\dagger \tilde{w}_1+\tilde{w}_2^\dagger \tilde{w}_2-\tilde{w}_1^\dagger \tilde{w}_2-\tilde{w}_2^\dagger \tilde{w}_1\right)\nonumber \\
&=\Tr(\rho_1)+\Tr(\rho_2)-\max_{\tilde{w}_1,\tilde{w}_2}\Tr(\tilde{w}_1^\dagger \tilde{w}_2+\tilde{w}_2^\dagger \tilde{w}_1) \nonumber \\
&=2-2\max_{\tilde{w}_1,\tilde{w}_2}\mathrm{Re}\left[\Tr(\tilde{w}_1^\dagger \tilde{w}_2)\right].
\end{align}
Since $\mathrm{Re}(x)\leq|x|$, it is clear that the condition for the maximum is to chose $\tilde{w}_1$, $\tilde{w}_2$, such that such $\tilde{w}_1^{\dagger}\tilde{w}_2$ is self-adjoint and positive definite
\begin{equation}\label{w1w2Real}
\tilde{w}_1^\dagger \tilde{w}_2=\tilde{w}_2^\dagger \tilde{w}_1>0.
\end{equation}

More explicit expressions may be obtained with the polar decomposition theorem. This theorem states that any operator $A$ (consider it to be finite dimensional for simplicity) can be decomposed as $A=|A|U_A$, where $|A|=\sqrt{A A^\dagger}$ and $U_A$ is a unitary operator \cite{Polar,rmp}. Furthermore for any unitary operator $U$ the following inequality holds
\begin{align}
\mathrm{Re}[\Tr(AU)]\leq|\Tr(AU)|&=|\Tr(|A|UU_A)|\nonumber \\
&=\left|\Tr\left(\sqrt{|A|}\sqrt{|A|}UU_A\right)\right|
\leq\sqrt{\left(\Tr|A|\right)\left[\Tr\left(U_A^\dagger U^\dagger |A| UU_A\right)\right]}=\Tr|A|
\end{align}
where in the penultimate step we have used the Cauchy-Schwarz inequality $|\Tr(A^\dagger B)|^2\leq\Tr(A^\dagger A)\Tr(B^\dagger B)$. Since the equality is reached for $U=U_A^\dagger$, we can write
\begin{equation}\label{maxU}
\max_U\mathrm{Re}[\Tr(AU)]=\Tr|A|.
\end{equation}
Now, the polar decomposition theorem on the amplitudes $\tilde{w}_1=\sqrt{\rho_1}U_1$ and $\tilde{w}_2=\sqrt{\rho_2}U_2$ jointly with Eq. \eqref{maxU} lead to
\begin{align}\label{maxU2}
\max_{\tilde{w}_1,\tilde{w}_2}\mathrm{Re}\left[\Tr(\tilde{w}_1^\dagger \tilde{w}_2)\right]&=\max_{U_1,U_2}\mathrm{Re}\left[\Tr(U_1^\dagger \sqrt{\rho_1}\sqrt{\rho_2}U_2)\right]\nonumber \\
&=\max_{U}\mathrm{Re}\left[\Tr(\sqrt{\rho_1}\sqrt{\rho_2}U)\right]=\Tr|\sqrt{\rho_1}\sqrt{\rho_2}|=\Tr\sqrt{\sqrt{\rho_1}\rho_2\sqrt{\rho_1}}.
\end{align}
Therefore, the Hilbert space distance between two parallel amplitudes $w_1$ and $w_2$ is equal to the Bures distance \cite{Bures} between $\rho_1$ and $\rho_2$,
\begin{equation}\label{parallelw}
\| w_1-w_2 \|^2=2-2\Tr\sqrt{\sqrt{\rho_1}\rho_2\sqrt{\rho_1}}, \quad \text{for } w_1\parallel w_2.
\end{equation}

Finally, note that the equality in \eqref{maxU2} is obtained for $U=U_{\sqrt{\rho_1}\sqrt{\rho_2}}^\dagger$, this is the adjoint of the unitary operator of the polar decomposition of $\sqrt{\rho_1}\sqrt{\rho_2}=|\sqrt{\rho_1}\sqrt{\rho_2}|U_{\sqrt{\rho_1}\sqrt{\rho_2}}$. Such a unitary operator is not uniquely defined unless $\sqrt{\rho_1}\sqrt{\rho_2}$ is full rank (and so $\sqrt{\rho_1}$, $\sqrt{\rho_2}$ and $|\sqrt{\rho_1}\sqrt{\rho_2}|$). In such a case, $U_{\sqrt{\rho_1}\sqrt{\rho_2}}$ is given by the equation
\begin{equation}\label{fullrank}
U_{\sqrt{\rho_1}\sqrt{\rho_2}}=|\sqrt{\rho_1}\sqrt{\rho_2}|^{-1}\sqrt{\rho_1}\sqrt{\rho_2}.
\end{equation}
Since from Eq. \eqref{maxU2} we have $U=U_2U_1^\dagger=U_{\sqrt{\rho_1}\sqrt{\rho_2}}^\dagger$, by using the unitarity of $U$ and Eq. \eqref{fullrank} we arrive at
\begin{equation}\label{U2U1}
U_2U_1^\dagger=\sqrt{\rho_2^{-1}}\sqrt{\rho_1^{-1}}\sqrt{\sqrt{\rho_1}\rho_2\sqrt{\rho_1}}.
\end{equation}
Hence, Eq. \eqref{U2U1} gives us the relation that the unitaries $U_1$ and $U_2$ must satisfy in order to minimize the distance between the associated amplitudes $w_1$ and $w_2$, i.e. in order to have $w_1 \parallel w_2$.

\subsection{Uhlmann Parallel Transport and Connection Form}

\begin{figure}[t]
\centering
\includegraphics[width=0.5\textwidth]{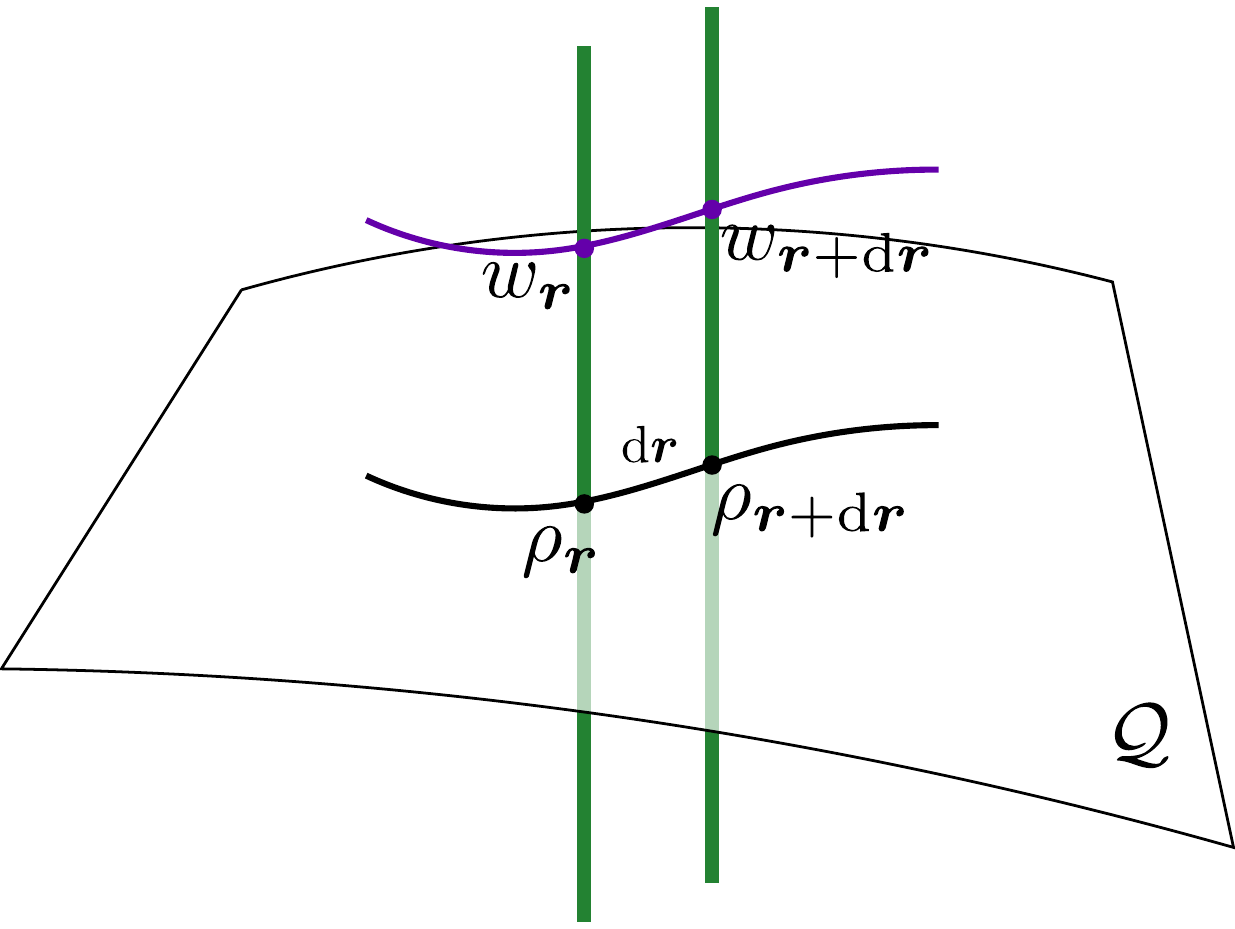}
\caption{Uhlmann parallel transport condition. The amplitude $w_{\bm{r}+{\rm d}\bm{r}}$ is the closest amplitude to $w_{\bm{r}}$ among the ones in the fiber at $\bm{r}+{\rm d}\bm{r}$.}
\label{plotParallelTransport}
\end{figure}

Before introducing the Uhlmann holonomy, we need to explain some of its ingredients: the Uhlmann parallel transport condition and the connection form. Consider a trajectory in the density matrix space $\mathcal{Q}$, $\rho_{\bm{r}}$. This induces a trajectory in the space $\mathcal{H}_w$, via the amplitudes $w_{\bm{r}}$, $\rho_{\bm{r}}=w_{\bm{r}}w^\dagger_{\bm{r}}$. Since for a point $\rho_{\bm{r}}$ of $\mathcal{Q}$ there are several amplitudes $w_{\bm{r}}$, a unique trajectory on $\mathcal{Q}$ induces multiple trajectories in $\mathcal{H}_w$ depending on which $w_{\bm{r}}$ are chosen. To select one of them, we define a parallel transport condition, i.e. for a point $\rho_{\bm{r}}$ with some amplitude $w_{\bm{r}}$, we consider the amplitude $w_{\bm{r}+{\rm d}\bm{r}}$ of the next point of the trajectory, $\rho_{\bm{r}+{\rm d}\bm{r}}$, to be the closest to $w_{\bm{r}}$ according to the Hilbert space distance, so $w_{\bm{r}}\parallel w_{\bm{r}+{\rm d}\bm{r}}$ in the sense of the previous section, see Fig. \ref{plotParallelTransport}. Such a path in $\mathcal{H}_w$ is known as a \emph{parallel lift} of path $\rho_{\bm{r}}$ in $\mathcal{Q}$. This is analogous to the Berry parallel transportation, where the U(1) phases of the consecutive state vectors $|\psi_{\bm{r}}\rangle$ and $|\psi_{\bm{r}+{\rm d}\bm{r}}\rangle$ are chosen such that the distance between them $\sqrt{(\langle\psi_{\bm{r}}|-\langle\psi_{\bm{r}+{\rm d}\bm{r}}|)(|\psi_{\bm{r}}\rangle-|\psi_{\bm{r}+{\rm d}\bm{r}}\rangle)}$ is minimal.

Let $\rho_{{\bm r}_0}$ be the initial state of the trajectory, under the square root section (or equivalently by the polar decomposition theorem) we have $w_{{\bm r}_0}=\sqrt{\rho_{{\bm r}_0}}U_{{\bm r}_0}$. For the sake of clarity we shall work for the moment on a gauge such that $U_{{\bm r}_0}=\mathds{1}$, and we shall refer to this gauge as the ``identity'' gauge. At some point $\bm{r}$ of the trajectory we have $w_{\bm{r}}=\sqrt{\rho_{\bm{r}}}V_{\bm{r}}$, where $V_{\bm{r}}$ is a unitary (Uhlmann holonomy)  constructed from the initial condition $w_{{\bm r}_0}=\sqrt{\rho_{{\bm r}_0}}$ by applying the parallel transport rule along the trajectory. The infinitesimal generator $A(\bm{r})$ of this unitary transformation $V_{\bm{r}}$ is called \emph{connection form} of the trajectory. Thus, if $\bm{r}(t)|_{t=0}^1$ is a parametrization of the trajectory $[\bm{r}(0)\equiv\bm{r}_0]$, $V_{\bm{r}}$ is fully determined by the differential problem
\begin{equation}\label{Adif}
\begin{cases}
\frac{dV_{\bm{r}(t)}}{dt}=A[\bm{r}(t)]V_{\bm{r}(t)},\\
V_{\bm{r}(0)}=\mathds{1},
\end{cases}
\end{equation}
whose formal solution can be written independently of the parametrization as
\begin{equation}\label{V}
V_{\bm{r}}=\mathcal{P}{\rm e}^{\int_{\bm{r}_0}^{\bm{r}} A(\bm{r}')d\bm{r}'},
\end{equation}
where $\mathcal{P}$ is the path ordering operator.

As we shall see, the Uhlmann connection $A$ is not uniquely defined unless the density matrix $\rho_{\bm{r}}$ is full rank. In order to compute $A$ in such a case, let us rewrite Eq. \eqref{Adif} in terms of differentials
\begin{equation}\label{Adif2}
{\rm d}V V^\dagger=A,
\end{equation}
where $A=\sum_iA_i{\rm d}r_i$ is understood as a differential one-form. Note that in Eq. \eqref{Adif}, $A[\bm{r}(t)]=\sum_iA_i\frac{dr_i}{dt}$. Then, consider some full rank density matrix $\rho_1=\rho$, and let displace it by an infinitesimal translation $\rho_2=\rho+{\rm d}\rho$. The basic condition to compute $A$ is that the corresponding infinitesimally closed amplitudes $w_1=\sqrt{\rho}V$ and $w_2=\sqrt{\rho+{\rm d}\rho}(V+{\rm d}V)$ fulfill  the parallel transport condition \eqref{parallelw}. Thus, Eq. \eqref{U2U1} leads to
\begin{equation} \label{VdVVdaga}
(V+{\rm d}V)V^\dagger=(\sqrt{\rho}+{\rm d}\sqrt{\rho})^{-1}\sqrt{\rho^{-1}}\sqrt{\sqrt{\rho}(\rho+{\rm d}\rho)\sqrt{\rho}}
\end{equation}
where we have used that for an infinitesimal translation ${\rm d}\rho$, $\sqrt{\rho+{\rm d}\rho}=\sqrt{\rho}+{\rm d}\sqrt{\rho}$. Following H\"ubner \cite{Hubner}, we introduce an auxiliary real parameter $s$ multiplying every differential:
\begin{equation}
(V+s{\rm d}V)V^\dagger=(\sqrt{\rho}+s{\rm d}\sqrt{\rho})^{-1}\sqrt{\rho^{-1}}\sqrt{\sqrt{\rho}(\rho+s{\rm d}\rho)\sqrt{\rho}}.
\end{equation}
The aim of this trick is to expand the above equation around $s=0$ and keep only the first order in $s$. In order words, we have that
\begin{equation}
\mathds{1}+s{\rm d}VV^\dagger=\mathds{1}+\frac{d}{ds}\left.\left[(\sqrt{\rho}+s{\rm d}\sqrt{\rho})^{-1}\sqrt{\rho^{-1}}\sqrt{\sqrt{\rho}(\rho+s{\rm d}\rho)\sqrt{\rho}}\right]\right|_{s=0}s,
\end{equation}
therefore
\begin{equation}\label{dA}
A={\rm d}VV^\dagger=\left.\left[\frac{d}{ds}(\sqrt{\rho}+s{\rm d}\sqrt{\rho})^{-1}\right]\right|_{s=0}\sqrt{\rho}+\rho^{-1}\left.\left[\frac{d}{ds}\sqrt{\sqrt{\rho}(\rho+s{\rm d}\rho)\sqrt{\rho}}\right]\right|_{s=0}.
\end{equation}
To give a closed expression for $A$, we determine its matrix elements with respect to the eigenbasis $\{|\psi_j\rangle\}$ of $\rho$, which we assume to be diagonalized with eigenvalues $\{p_j\}$. For the first term in \eqref{dA} we have
\begin{align}
(\sqrt{\rho}+s{\rm d}\sqrt{\rho})^{-1}\sqrt{\rho}&=\left[\sqrt{\rho}(\mathds{1}+s\sqrt{\rho^{-1}}{\rm d}\sqrt{\rho})\right]^{-1}\sqrt{\rho}\nonumber\\
&=(\mathds{1}+s\sqrt{\rho^{-1}}{\rm d}\sqrt{\rho})^{-1}\sqrt{\rho^{-1}}\sqrt{\rho}=\mathds{1}-s\sqrt{\rho^{-1}}{\rm d}\sqrt{\rho},
\end{align}
where in the last step the inverse has been taken at first order in $s$. Hence
\begin{equation}
\left.\left[\frac{d}{ds}(\sqrt{\rho}+s{\rm d}\sqrt{\rho})^{-1}\right]\right|_{s=0}\sqrt{\rho}=-\sqrt{\rho^{-1}}{\rm d}\sqrt{\rho},
\end{equation}
and in the eigenbasis of $\rho$,
\begin{equation}
-\langle\psi_i|\sqrt{\rho^{-1}}{\rm d}\sqrt{\rho}|\psi_j\rangle=-\frac{1}{\sqrt{p_i}}\langle\psi_i|{\rm d}\sqrt{\rho}|\psi_j\rangle.
\end{equation}

The second term of \eqref{dA} is a bit more involved. Define $K(s):=\sqrt{\sqrt{\rho}(\rho+s{\rm d}\rho)\sqrt{\rho}}$; we need to compute $\rho^{-1}K'(0)$. By differentiating the square,
\begin{equation}
\{[K(s)]^2\}'=K'(s)K(s)+K(s)K'(s)=\sqrt{\rho}{\rm d}\rho\sqrt{\rho},
\end{equation}
as $K(0)=\rho$ we have
\begin{equation}
K'(0)\rho+\rho K'(0)=\sqrt{\rho}{\rm d}\rho\sqrt{\rho}.
\end{equation}
After taking matrix elements in the eigenbasis of $\rho$,
\begin{equation}
(p_i+p_j)\langle\psi_i|K'(0)|\psi_j\rangle=\sqrt{p_ip_j}\langle\psi_i|{\rm d}\rho|\psi_j\rangle.
\end{equation}
Thus, the matrix elements of $\rho^{-1}K'(0)$ are
\begin{equation}
\langle\psi_i|\rho^{-1}K'(0)|\psi_j\rangle=\frac{1}{p_i}\langle\psi_i|K'(0)|\psi_j\rangle=\frac{\sqrt{p_j}}{\sqrt{p_i}(p_i+p_j)}\langle\psi_i|{\rm d}\rho|\psi_j\rangle.
\end{equation}
Now, since ${\rm d}\rho={\rm d}(\sqrt{\rho}\sqrt{\rho})=({\rm d}\sqrt{\rho})\sqrt{\rho}+\sqrt{\rho}({\rm d}\sqrt{\rho})$, we obtain
\begin{equation}
\langle\psi_i|\rho^{-1}K'(0)|\psi_j\rangle=\frac{\sqrt{p_j}(\sqrt{p_i}+\sqrt{p_j})}{\sqrt{p_i}(p_i+p_j)}\langle\psi_i|{\rm d}\sqrt{\rho}|\psi_j\rangle
\end{equation}
and finally the matrix elements of the connection form,
\begin{equation}
\langle\psi_i|A|\psi_j\rangle=\left(-\frac{1}{\sqrt{p_i}}+\frac{\sqrt{p_j}(\sqrt{p_i}+\sqrt{p_j})}{\sqrt{p_i}(p_i+p_j)}\right)\langle\psi_i|{\rm d}\sqrt{\rho}|\psi_j\rangle=\frac{\left(\sqrt{p_j}-\sqrt{p_i}\right)}{(p_i+p_j)}\langle\psi_i|{\rm d}\sqrt{\rho}|\psi_j\rangle.
\end{equation}
This can be rewritten in terms of a commutator,
\begin{equation}\label{dAfinal}
A=\sum_{i,j}|\psi_i\rangle\frac{\langle\psi_i|[{\rm d}\sqrt{\rho},\sqrt{\rho}]|\psi_j\rangle}{(p_i+p_j)}\langle\psi_j|.
\end{equation}
Note that this connection form has only zeroes on its diagonal and is skew-adjoint, so that the Uhlmann connection is special unitary.

For the particular case of $2\times2$ density matrices, because of the normalization condition $p_1+p_2=1$, Eq. \eqref{dAfinal} can be written as
$A=[{\rm d}\sqrt{\rho},\sqrt{\rho}]$. This formula was also obtained by Dittmann and Rudolph \cite{Dittmann}. Moreover, expanding in terms of Pauli matrices $\sqrt{\rho}=a_0+\bm{a}\cdot\bm{\sigma}$, we may write $A=2{\rm i} ({\rm d}\bm{a}\times\bm{a})\cdot \bm{\sigma}$ \cite{Hubner}.

The result \eqref{dAfinal} has been obtained under the requirement that $\rho$ is full rank. If such is not the case, for $w_1=w_{\bm{r}}$ the formula \eqref{parallelw} can be fulfilled for several $w_2=w_{\bm{r}+{\rm d}\bm{r}}$ and there is not a unique solution and so a unique connection. One of these solutions can be obtained from continuity if we have a family of full rank states $\rho_{\lambda}$ which approaches the desired non-full rank state as $\lambda\rightarrow0$. Then a connection can be obtained with the formula \eqref{dAfinal} for $\rho_{\lambda}$, taking the limit $\lambda\rightarrow0$ at the end of the computation. We will come back to this point at the end of the following section.

Once a connection form $A$ is defined, we can construct its associate curvature form via the standard relation,
\begin{equation}
F={\rm d}A+A\wedge A,
\end{equation}
where $\wedge$ denotes the exterior product and the action of ${\rm d}$ is understood as the exterior derivative \cite{Nakahara}. For the sake of clarity, let us write this equation in components. Expanding  $A=\sum_i A_i {\rm d}r_i$, we find the usual expression of the field strength,
\begin{equation}\label{FnonAbelian}
F_{ij}=\partial_i A_j-\partial_j A_i +[A_i,A_j],
\end{equation}
where $F=\frac{1}{2}\sum_{i,j} F_{ij} {\rm d}r_i\wedge{\rm d}r_j$.

This curvature form $F$ allows us to construct the Chern classes of the Uhlmann fiber bundle, which is the standard way to obtain topological invariants. We shall come back to this point in Sec.\ref{sec:IV}A.

\subsection{Gauge Transformations}

In the previous subsection we obtained an expression for the Uhlmann connection in the ``identity'' gauge, i.e. for the trajectory $\rho_{\bm r}$, we take initially $w_{{\bm r}(0)}=\sqrt{\rho_{{\bm r}(0)}}$ and construct the next point of the trajectory in $\mathcal{H}_w$ applying the Uhlmann parallel transport rule, such that $w_{\bm r}=\sqrt{\rho_{\bm r}}V_{\bm r}$ where $V_{\bm r}$ is the Uhlmann holonomy of the trajectory. Now consider a change of gauge by left-multiplying the amplitudes by some $\bm{r}$-dependent unitary $U_{\bm{r}}$, $w_{\bm{r}_0}=\sqrt{\rho_{\bm{r}_0}}U_{\bm{r}_0}$, so that $w_{\bm r}=\sqrt{\rho_{\bm r}}U_{{\bm r}}V_{U,{\bm r}}$, where $V_{U,{\bm r}}$ is the Uhlmann holonomy in the new gauge. To compute $V_{U,{\bm r}}$, we first obtain the connection form in the new gauge.

Similarly to the previous section we have to impose that two infinitesimally closed amplitudes are parallel. Under an infinitesimal displacement $w_1=\sqrt{\rho}UV_{U}$ changes to $w_2=\sqrt{\rho+{\rm d}\rho}(U+{\rm d}U)(V_{U}+{\rm d}V_{U})$, so that according to Eq. \eqref{U2U1}, $w_1\Vert w_2$ if
\begin{equation}
(U+{\rm d}U)(V_U+{\rm d}V_U)V_{U}^\dagger U^\dagger=(\sqrt{\rho}+{\rm d}\sqrt{\rho})^{-1}\sqrt{\rho^{-1}}\sqrt{\sqrt{\rho}(\rho+{\rm d}\rho)\sqrt{\rho}}.
\end{equation}
From Eq. \eqref{VdVVdaga} and subsequent ones, the right hand side is simply $\mathds{1}+A_{\mathds{1}}$, where $A_{\mathds{1}}$ denotes the connection in the ``identity'' gauge. Moreover, the left hand side at first order in differentials is $\mathds{1}+{\rm d}UU^\dagger+U{\rm d}V_UV_U^\dagger U^\dagger$, so we obtain,
\begin{equation}
{\rm d}UU^\dagger+U{\rm d}V_UV_U^\dagger U^\dagger=A_{\mathds{1}}.
\end{equation}
Since by definition, $A_U={\rm d}V_UV_U^\dagger$ is the connection form in the new gauge, we find the transformation rule
\begin{equation}\label{Agauge}
A_U= U^\dagger A_{\mathds{1}} U- U^\dagger{\rm d}U.
\end{equation}
In general, under a change of gauge $U_2=U_1U$ $(w\rightarrow wU)$, we straightforwardly obtain the transformation law
\begin{equation}
A_{U_2}= U^\dagger A_{U_1} U- U^\dagger{\rm d}U.
\end{equation}
Now we can calculate the holonomy in the new gauge,
\begin{equation}\label{VUpath}
V_{U,\bm{r}}=\mathcal{P}{\rm e}^{\int_{\bm{r}_0}^{\bm{r}} A_U(\bm{r}')d\bm{r}'}.
\end{equation}
To that aim, consider the parametrization $\bm{r}(t)|_{t=0}^1$ and compute the $t$-derivative of $U[\bm{r}(t)]V_{U,\bm{r}(t)}U^\dagger[\bm{r}(0)]$:
\begin{align}
\frac{d}{dt}\left\{U[\bm{r}(t)]V_{U,\bm{r}(t)}U^\dagger[\bm{r}(0)]\right\}&=\frac{dU[\bm{r}(t)]}{dt}V_{U,\bm{r}(t)}U^\dagger[\bm{r}(0)]+U[\bm{r}(t)]\frac{dV_{U,\bm{r}(t)}}{dt}U[\bm{r}(0)]\\
&=\frac{dU[\bm{r}(t)]}{dt}V_{U,\bm{r}(t)}U^\dagger[\bm{r}(0)]+U[\bm{r}(t)]A_U[\bm{r}(t)] V_{U,\bm{r}(t)}U[\bm{r}(0)]\\
&=A_{\mathds{1}}[\bm{r}(t)] U[\bm{r}(t)]V_{U,\bm{r}(t)}U[\bm{r}(0)],
\end{align}
where we have used the definition Eq. \eqref{VUpath} and introduced the gauge transformation Eq. \eqref{Agauge}. Since  $U[\bm{r}(t)]V_{U,\bm{r}(t)}U^\dagger[\bm{r}(0)]|_{t=0}=\mathds{1}$; $\{U[\bm{r}(t)]V_{U,\bm{r}(t)}U^\dagger[r(0)]\}$ satisfies the same differential problem as $V_{\mathds{1},\bm{r}(t)}$, Eq. \eqref{Adif}, and the uniqueness of the solution implies that:
\begin{equation}\label{Vgauge}
V_{\mathds{1},\bm{r}(t)}=U[\bm{r}(t)]V_{U,\bm{r}(t)}U^\dagger[\bm{r}(0)] \Rightarrow V_{U,\bm{r}}=U^\dagger(\bm{r})V_{\mathds{1},\bm{r}}U(\bm{r}_0).
\end{equation}

Therefore, the Uhlmann holonomy of the trajectory is clearly gauge covariant as expected. This is a fundamental requirement in order to define gauge invariant quantities out of this construction.

\subsection{Uhlmann Geometric phase}

Once the concept of Uhlmann holonomy has been presented, it is interesting and convenient to define gauge invariant quantities associated to it. These are constructed out of the relative phases of the amplitudes that have been parallelly transported. Actually, this is the most important part of the analysis, since we want to relate the mathematical concepts to physical observables that are represented by gauge invariants in the theory.

First of all, consider some trajectory on $\mathcal{Q}$, with initial and final points $\rho_0$ and $\rho_1$ respectively. For some initial amplitude $w_0$, $\rho_0=w_0w^\dagger_0$, a parallel lift leads to some amplitude $w_1$, $\rho_1=w_1w^\dagger_1$. The \emph{Uhlmann geometric phase} is defined as
\begin{equation}
\Phi_{\rm U}:=\arg(w_0,w_1)=\arg \Tr(w^\dagger_0w_1),
\end{equation}
which is a gauge-independent quantity.  Indeed, in the identity gauge, $w_0=\sqrt{\rho_0}$ and $w_1=\sqrt{\rho_1}$ we have
\begin{equation}
\Phi_{\rm U}=\arg \Tr(\sqrt{\rho_0} \sqrt{\rho_1}V).
\end{equation}
Under a gauge transformation $w_0\rightarrow w_0U_0$, $w_1\rightarrow w_1U_1$ and according to Eq. \eqref{Vgauge}, $V\rightarrow U^\dagger_1 V U_0$, so  $\Phi_{\rm U}\rightarrow\arg \Tr(U^\dagger_0 \sqrt{\rho_0} \sqrt{\rho_1} U_1 U_1^\dagger VU_0)=\Phi_{\rm U}$. Therefore $\Phi_{\rm U}$ is gauge invariant.

To better visualize the  geometric meaning  of $\Phi_{\rm U}$, let us consider the trajectory on $Q$ to be closed. Then $\rho_0=\rho_1=\rho$ and we have
\begin{equation}\label{phiU}
\Phi_{\rm U}=\arg \Tr(\rho V)=\arg \Tr\left(\rho \mathcal{P}{\rm e}^{\oint A}\right).
\end{equation}

\begin{figure}[t]
\centering
\includegraphics[width=0.8\textwidth]{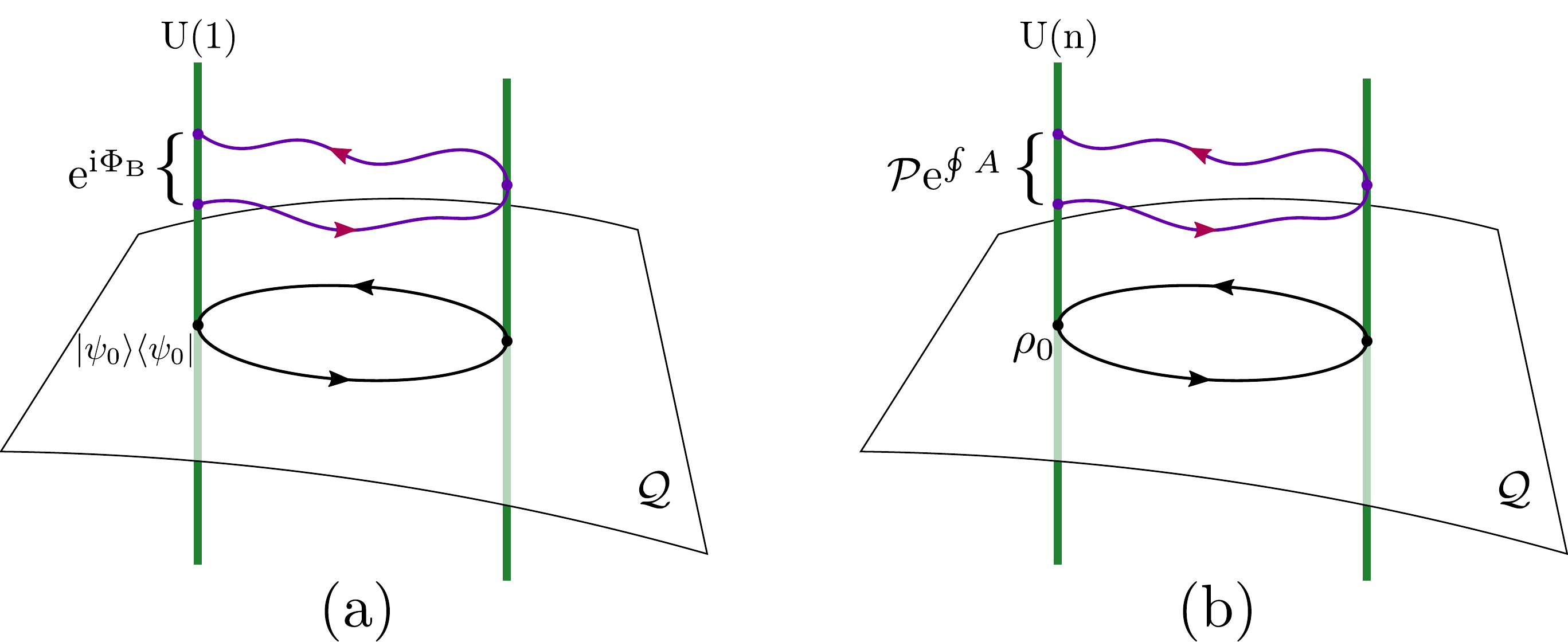}
\caption{Comparison between Berry and Uhlmann Holonomies. In the approach of Berry, valid for pure states (a), after a closed loop on the space $\mathcal{Q}$, the final state vector differs by a phase $\Phi_{\rm B}$ with respect to the initial state vector. In the case of Uhlmann, formulated for density matrices (b), the difference between initial and final amplitude is, in general, the unitary matrix given by $\mathcal{P}{\rm e}^{\oint A}$.}
\label{plotHolonomies}
\end{figure}

In the Uhlmann approach, after a closed trajectory in $\mathcal{Q}$ the amplitude gains an extra geometric ${\rm U}(n)-$phase, $\mathcal{P}{\rm e}^{\oint A}$. This is the equivalent to the Berry geometric ${\rm U}(1)-$phase in this non-abelian approach, see Fig. \ref{plotHolonomies}. The equation \eqref{phiU} defines a real gauge invariant quantity extracting the non-abelian ${\rm U}(n)-$phase which can be understood as the relative phase between $w_0$ and $w_1$. Basically, this is because by using the amplitude-purification isomorphism Eq. \eqref{isomorfismo}, $\Phi_U$ can be equivalently written as
\begin{equation}
\Phi_{\rm U}=\arg \langle w_0|w_1\rangle.
\end{equation}
Actually, this interpretation has suggested some schemes to measure Uhlmann geometric phases \cite{medidaU}.

Another important concept needs to be introduced in order to understand the properties of the Uhlmann phase. This is the difference between an \emph{absolute holonomy} and a \emph{pointed holonomy}.
The mapping
\begin{equation}
M_U:~~w \longrightarrow V,
\end{equation}
is a so-called pointed holonomy. This means, that even if the trajectory in $\mathcal{Q}$ is closed, in general the holonomy depends on the initial point of the path. Nonetheless, we have identified cases in which the pointed holonomy reduces to an \emph{absolute holonomy} becoming independent of the initial point \cite{Viyuela_et_al14,Viyuela_et_al14_2D}. In addition, it is always true that the mapping
\begin{equation}
\tilde{M}_{\rm U}:~~w \longrightarrow {\rm spect}(V),
\end{equation}
where ${\rm spect}(V)$ stands for the spectrum of the unitary $V$, is an absolute holonomy.

To conclude this section, consider the case of a pure state trajectory $\rho_{\bm{r}}=|\psi_{\bm{r}}\rangle\langle\psi_{\bm{r}}|$. As we have already pointed, the Uhlmann connection in this case is not uniquely defined because the state $\rho_{\bm{r}}$ is not full rank. There are two approaches to overcome this difficulty:

\begin{itemize}
\item As aforementioned, a first approach to compute the Uhlmann phase factor in this situation is to consider the connection $A^{\lambda}$ computed via \eqref{dAfinal} for a full rank matrix trajectory $\rho^{\lambda}_{\bm{r}}$ such that $\lim_{\lambda\rightarrow0}\rho^{\lambda}_{\bm{r}}=|\psi_{\bm{r}}\rangle\langle\psi_{\bm{r}}|$. Thus, the Uhlmann phase factor is obtained by formula \eqref{phiU} with $V=\lim_{\lambda\rightarrow0}\mathcal{P}{\rm e}^{\oint A^{\lambda}}$. This is a typical approach when $|\psi_{\bm{r}}\rangle$ is seen as the ground state of some Hamiltonian and we want to study temperature effects. Then, the family of full rank states will be the Gibbs states which approach the ground state in the zero temperature limit.

\item An alternative approach is to look for another solution to the optimization problem \eqref{parallelw}. For example, consider two pure states, $|\psi_1\rangle$ and $|\psi_2\rangle$, the optimization problem \eqref{parallelw} can be easily solved in a gauge with amplitudes of the form of $w_1=|\psi_1\rangle\langle a| {\rm e}^{{\rm i}\phi_1}$, $w_2=|\psi_2\rangle\langle a| {\rm e}^{{\rm i}\phi_2}$ where $|a\rangle$ is some arbitrary $\bm{r}$-independent vector, and the phases are chosen such that $(\phi_1-\phi_2)=\arg\langle\psi_1|\psi_2\rangle$. This can be directly checked in \eqref{parallelw} and also derived from condition \eqref{w1w2Real}. Thus, in this gauge we can satisfy the parallel transport rule just by varying a U(1) factor, ${\rm e}^{{\rm i}\phi}$, extracted from the general U(n) gauge freedom. More concretely, we associate the amplitudes $w_{\bm{r}}=|\psi_{\bm{r}}\rangle\langle a|{\rm e}^{{\rm i }\phi(\bm{r})}$ with the trajectory $\rho_{\bm{r}}=|\psi_{\bm{r}}\rangle\langle\psi_{\bm{r}}|$. Then we have
\begin{align}
w_{\bm{r}}^\dagger (w_{\bm{r}}+{\rm d}w_{\bm{r}})&=|a\rangle\langle a|+{\rm e}^{-{\rm i }\phi(\bm{r})} |a\rangle\langle \psi_{\bm{r}}| {\rm d}\left(|\psi_{\bm{r}}\rangle\langle a|{\rm e}^{{\rm i }\phi(\bm{r})}\right)\nonumber\\
&=\left[1+\langle \psi_{\bm{r}}| {\rm d}\psi_{\bm{r}}\rangle+{\rm i d}\phi(\bm{r})\right]|a\rangle\langle a|,
\end{align}
where due to normalization, $\langle \psi_{\bm{r}}| {\rm d}\psi_{\bm{r}}\rangle$ is purely imaginary. The parallel transport law in the form of $\eqref{w1w2Real}$, with $w_1=w_{\bm{r}}$ and $w_2=w_{\bm{r}}+{\rm d}w_{\bm{r}}$ leads to the condition
\begin{equation}\label{dphi}
{\rm d}\phi(\bm{r})={\rm i}\langle \psi_{\bm{r}}| {\rm d}\psi_{\bm{r}}\rangle.
\end{equation}
If we denote $|\psi_{\bm{r}_0}\rangle\langle \psi_{\bm{r}_0}|$ the initial and final state, we have $w_0=|\psi_{\bm{r}_0}\rangle\langle a|{\rm e}^{{\rm i }\phi(\bm{r}_0)}$, and the parallel transported final amplitude will be $w_1=|\psi_{\bm{r}_0}\rangle\langle a|{\rm e}^{\oint {\rm d}\phi(\bm{r})}{\rm e}^{{\rm i }\phi(\bm{r}_0)}$, where ${\rm e}^{\oint {\rm d}\phi(\bm{r})}$ is the accumulated phase after the parallel translation. Thus, by using Eq. \eqref{dphi} the Uhlmann phase factor $\Phi_{\rm U}=\arg \Tr(w^\dagger_0w_1)$ yields
\begin{equation}
\Phi_{\rm U}=\arg\Tr\left[{\rm e}^{-{\rm i }\phi(\bm{r}_0)} |a\rangle\langle \psi_{\bm{r}_0}|\psi_{\bm{r}_0}\rangle\langle a|{\rm e}^{\oint {\rm d}\phi(\bm{r})}  {\rm e}^{{\rm i }\phi(\bm{r}_0)}\right]={\rm i}\oint\langle \psi_{\bm{r}}| {\rm d}\psi_{\bm{r}}\rangle \quad {\rm mod}\ 2\pi,
\end{equation}
which is nothing but the usual Berry geometric phase.
\end{itemize}

Very importantly, since the Uhlmann phase is independent of the gauge, the two approaches must provide the same result. This is going to be crucial in the next section in order to obtain consistent results when taking the limits of zero temperature and pure states.

\section{DENSITY-MATRIX TOPOLOGICAL INSULATORS AND SUPERCONDUCTORS: TOPOLOGICAL UHLMANN PHASE}
\label{sec:III}

In this section we will describe how the Uhlmann geometric phase acquires a topological character in certain 1D condensed matter systems. The problem of combining the study of topological phases with the fact of having mixed states and finite temperature is a very relevant and had remained vaguely explored for a long time. In a set of recent papers \cite{Viyuela_et_al12,Rivas_et_al13,Viyuela_et_al14,Viyuela_et_al14_2D}, we tried to deal with this issue and proposed some methods to generalize the notion of SPT order to finite temperature and to systems with dissipation.

The most characteristic feature of a geometric phase, as the Berry phase, is the fact that it does not depend on properties such as the gauge or the speed at which a given path is traversed but only on the geometry of the path. In the case of closed paths, this is manifested by the fact that the geometric phase only depends on the area enclosed by the trajectory on the manifold defined as parameter space.

There are cases, however, where the path covered during the holonomy coincides with one of the topologically non-trivial paths of a given manifold. In particular, in condensed matter, we know that due to translational invariance, the \emph{crystalline momentum} ${\bf k}$ defines the so-called Brillouin Zone (BZ) and characterizes the eigenstates of the system. Then the phase difference picked up by the system when parallelly transported from an initial momentum ${\bf k}_0$ to a final one ${\bf k}_0+{\bf G}$, where ${\bf G}$ is a reciprocal lattice vector, is gauge invariant and a geometric phase too. This was proven by Zak \cite{Zak_89} and the arguments work both in 1D and in higher dimensional systems. This particular Berry phase is also called Zak phase. Actually, in 1D systems, Zak \cite{Zak_89} showed that the Zak phase is quantized and equal to $0$ or $\pi$ in the presence of inversion symmetry.

On the other hand, the Uhlmann phase was initially proposed as a generalization of the Berry phase to density matrices, but it can also be applied to condensed matter systems in the spirit of the Zak phase. Actually, this is very fundamental since it deals with the problem of generalizing the notion of topology to systems with dissipation and finite temperature. Indeed, we studied this problem in detail \cite{Viyuela_et_al14} and found that when applied to representative models for topological insulators and superconductors, the Uhlmann phase acquired a topological character and is also quantized to $0$ or $\pi$ depending on whether the system is in a topological phase or not.

Specifically, we consider 1D Hamiltonians with two bands in momentum space. We denote by $a$ and $b$ two different types of fermionic band operators respectively, in such a way that in the spinor representation of $\psi_k=(a_k,b_k)^{\rm t}$ for insulators, while for superconductors, $\psi_{k}=(c_{k},c^{\dagger}_{-k})^{\rm t}$ is the Nambu spinor for paired fermions with opposite crystalline momentum \cite{LibroBernevig}. We take lattice spacing $a=1$ throughout the text. The Hamiltonian of these systems can be written as
\begin{equation}\label{H}
H=\sum_k\psi_k^{\dagger}H_k\psi_k,
\end{equation}
where $H$ is a quadratic form on $\psi$ and $H_k$ is a $2\times2$ matrix. We aim at studying the properties of this Hamiltonian under a displacement in $k$ along a circuit which encloses the whole Brillouin zone, which will be the circle $\mathit{S}^1$.

Note that $H_k$ can be decomposed in the following form
\begin{equation}\label{Hk}
H_k=f(k)\mathds{1} +\frac{\Delta_k}{2}{\bm n}_k\cdot{\bm \sigma}.
\end{equation}
Here, $\bm{\sigma}=(\sigma_x,\sigma_y,\sigma_z)$ are the Pauli matrices, $\Delta_k$ corresponds to the gap of $H_k$ and $f(k)$ denotes some function of $k$. The unit vector ${\bm n}_k=(\sin{\theta}\cos{\phi},\sin{\theta}\sin{\phi},\cos{\theta})$ is called \emph{winding vector}, where $\theta$ and $\phi$ are $k-$dependent spherical coordinates.

For a thermalization process preserving particle number, with the Fermi energy set in the middle of the gap, the connection in Eq. \eqref{dAfinal} associated to this equilibrium (thermal) state, $\rho_k^\beta={\rm e}^{-\beta H_k}/\Tr\left({\rm e}^{-\beta H_k}\right)$
where $T=1/\beta$ denotes temperature, is
\begin{equation}
A=m^k_{12}\langle u_{-}^k\ket{\partial_k u_{+}^k}\ket{u_{-}^k}\bra{u_{+}^k}dk+\text{h.c.}
\label{Arho2}
\end{equation}
where $m^k_{12}:=1-{\rm sech}\left(\frac{\Delta_k}{2T}\right)$.

Moreover, making use of the discrete symmetries \cite{Ludwig,Kitaev_2009} that restrict the movement of the winding vector ${\bm n}_k$ as a function of $k$, only two components ${n}^i_k$ and ${n}^j_k$ with $i\not=j$ are non-zero.
In order to obtain a simplified expression for the Uhlmann phase, we fix the gauge of the eigenvectors of $H_k$ in Eq. \eqref{Hk} in such a way that the off-diagonal overlap $\langle u_{-}^k\ket{\partial_k u_{+}^k}$ and the winding vector components are related in this manner
\begin{equation}
\langle u_{-}^k\ket{\partial_k u_{+}^k}=\frac{\partial_k {n}_k^i}{2n_k^j}.
\label{overlap}
\end{equation}
Note that as Eq. \eqref{phiU} is gauge invariant, we can fix the most convenient gauge to perform the calculation of the Uhlmann phase.

Thus, using Eq. \eqref{Arho2} with this particular gauge choice and simplifying Eq. \eqref{phiU}, we obtain a more simplified expression
\begin{equation}
\Phi_{\rm U}=\arg\Bigg\{\cos(\pi\omega_1)\cos\bigg[\oint\bigg(\frac{\partial_k {n}_k^i}{2n_k^j}\bigg)\text{sech}\left(\frac{\Delta_k}{2T}\right)dk\bigg]\Bigg\}.
\label{phiU2}
\end{equation}
where $\omega_1:=\frac{1}{2\pi}\oint\left(\frac{\partial_k {n}_k^i}{n_k^j}\right)dk$ is the winding number of ${\bm n}_k$ \cite{Viyuela_et_al14}, a topological indicator characterizing the mapping $\mathit{S}^1\longrightarrow \mathit{S}^1$, from $k~\longrightarrow~{\bm n}_k$.

In the case of being in a trivial phase with $\omega_1=0$, the Uhlmann phase is always zero. However, for nontrivial topological regions $\omega_1=\pm1$, we can obtain either $\Phi_{\rm U}=0$ or $\Phi_{\rm U}=\pi$ depending on the value of $T$. Thus, the topological order as accounted by $\Phi_{\rm U}$ can be also non-trivial for a range of temperatures $T<T_c$. 
In \cite{Viyuela_et_al14} we computed $\Phi_{\rm U}$  finite temperature for the three representative models models of topological insulators and superconductors, showing this critical effect with temperature represented in different phase diagrams.

\begin{figure}[t]
\begin{center}
\includegraphics[width=0.8\textwidth]{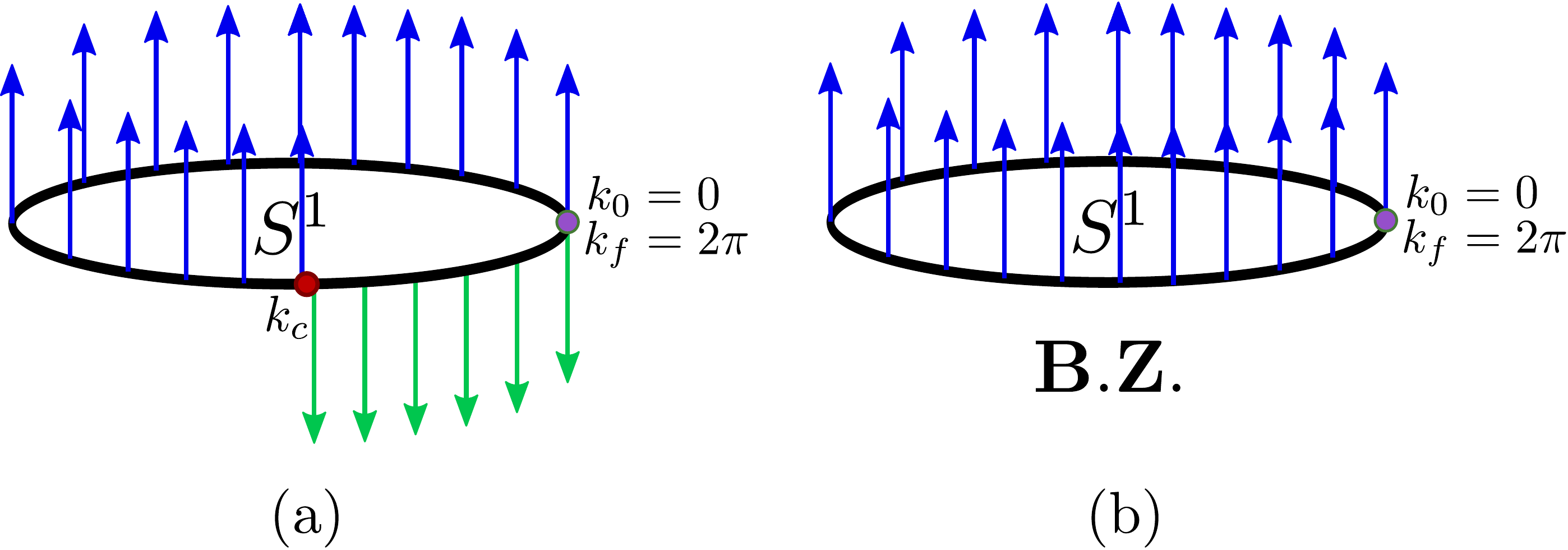}
\end{center}
\caption{Behavior of the Uhlmann phase during the parallel transport from $k_0=0$ to $k_f=2\pi$. We identify two situations: (a) Non-trivial topology highlighted by the presence of a topological kink at $k_c$. (b) Trivial order where the topological kink disappears. A detailed explanation of the figure can be found in the main text.}
\label{kink}
\end{figure}

On the other hand, the appearance of a critical temperature $T_c$ in the Uhlmann phase can be interpreted geometrically in a pictorial and intuitive way.
First of all, for the sake of illustration, we can visualize the amplitudes (or purifications) as arrows with fixed length. The angle between two arrows will be given by the relative phase between the corresponding amplitudes. In Fig. \ref{kink}, we sketch different situations where the amplitudes (arrows) are transported according to the Uhlmann's parallel condition along a closed loop that covers the whole Brilluoin zone $S^1$. Here we present a summary of the different behaviors:
\begin{enumerate}
\item At $T=0$ two things may happen. If the system is in a trivial phase, then the arrow is transported pointing always the same direction, which means that the relative phase between the amplitudes does not change. The initial and final arrows are parallel as in Fig. \ref{kink}(b), so that $\Phi_{\rm U}^{0}=\Phi_{\rm B}=0$. In the case that the system is in a topological phase, the Uhlmann phase remains 0 (the arrows are parallel) up to a certain critical point $k_c$, when the direction of the arrows is reversed and hence the relative phase between amplitudes is equal to $\pi$. This constitutes a \emph{topological kink}. From there, the arrow does not flip anymore until the final point of the path $k_f=2\pi$ is reached, as in Fig. \ref{kink}(a). Thus, the phase $\Phi_{\rm U}^0=\Phi_{\rm B}=\pi$.

\item At $T\not=0$, we can identify three different possibilities. If the system was in a trivial phase at $T=0$ then $\Phi_{\rm U}=0,~\forall T$, as we have already proven, being again in the case of Fig. \ref{kink}(b). If at $T=0$ the system was in a topological phase, then at finite temperature there are two options. Whenever $T<T_c$, the situation in Fig. \ref{kink}(a) is reproduced, the Uhlmann phase remains 0 up to the critical point which is a function of $T$. From that point, the arrow is reversed and the Uhlmann phase is equal to $\pi$ until the close path is completed. An increase of temperature shifts the position of $k_c$ towards the end of the Brillouin zone. However, if $T>T_c$, then we recover again Fig. \ref{kink}(b) with no phase change $\Phi_{\rm U}=0$. The position of $k_c$ has already reached the end of the Brillouin zone and the topological kink can no longer occur along the path between $k_0=0$ and $k_f=2\pi$.
\end{enumerate}

To conclude, we have derived a connection between the existence of a critical temperature $T_c$ and the presence of a topological kink structure \cite{Coleman} in the Uhlmann holonomy. Actually, the same relation can be stablished with the parameters of the Hamiltonian $H$, that drives the system out of the topological phase. This is because the Uhlmann phase places on equal footing $T$ and these coupling constants. The reason to call this kink topological is actually well established. The critical momentum $k_c$ divides the holonomy in two different topological sectors according to the value of the final point of the path $k_f$. The first sector covers the region with $k_f<k_c$ and $\Phi_{\rm U}=0$, and the second sector starts when $k_f>k_c$ and $\Phi_{\rm U}=\pi$. It is clear that these two different sections cannot be smoothly connected without crossing the singular point $k_f=k_c$. In the trivial topological regime $\Phi_{\rm U}=0$ however, we can associate every point in the sector with the same class. To be more specific, $\Phi_{\rm U}$ is invariant under changes of the final point of the holonomy $k_f\in(0,k_c)$. The same happens within the other non-trivial topological sector with $\Phi_{\rm U}=\pi$ once we consider $k_f\in(k_c,2\pi)$. The effect of the temperature in the Uhlmann parallel transport can be understood as a displacement of the topological kink along $k$-space. When the temperature reaches the critical value $T_c$, the critical momentum $k_c$ has been pushed out to the edge of the Brillouin zone and the topological kink disappears. Complementary, the critical temperature $T_c$ can be viewed as the maximum amount of noise/disorder such that the Uhlmann holonomy along the Brilluoin zone displays a topological kink structure.

\section{2D UHLMANN TOPOLOGICAL INVARIANTS}
\label{sec:IV}

In the previous section, we have shown that in 1D condensed matter systems, the Uhlmann geometric phase acquires a topological character when applied to topological insulators and superconductors. Actually, the relation of topological phases for pure states with invariants of SPT order in more than 1D has been extensively studied, in particular see \cite{Resta,Xiao}. Nonetheless, the concept of a geometric phase or a topological phase is not at all restricted to one-dimension and can be defined in higher-dimensional spaces.

The problem of generalizing the Uhlmann phase formalism for 2D cases is not straightforward at all  \cite{Viyuela_et_al14_2D, Arovas14}. This has to do with the triviality of the Chern classes associated with the Uhlmann connection of non-singular density matrices, as was first proven in Ref. \cite{Viyuela_et_al14}. Different but equivalent ways to solve this problem were proposed in Ref. \cite{Viyuela_et_al14_2D, Arovas14}. In Ref. \cite{Arovas14} an analysis of the spectrum of the whole holonomy matrix was carried out, while in Ref. \cite{Viyuela_et_al14_2D} a unique topological invariant called the Uhlmann number was first proposed to characterize the topological order of two-band topological insulators and superconductors at finite temperature in 2D.

In this section we review the theory of topological phases for mixed states in 2D, and give a detailed explanation of a particularly useful way to define topological invariants inside the Uhlmann formalism. As a concrete example, we will characterize the topological phase diagram of the QHZ model for a topological insulator at finite temperature.

\subsection{Triviality of the Uhlmann Chern classes}

At zero temperature, there is a standard way to define topological invariants in two-dimensional symmetry protected topological order systems out of Chern numbers. In this case, a fiber bundle is formed out from the pure states eigenvectors $|u_{\bm{k}}\rangle$ corresponding to the lower energy band, and the fiber is linked to the ${\rm U}(1)$ gauge freedom that each eigenvector has. As aforementioned, the first Chern number associated to the first Chern class of fiber bundle can be related to the transverse or quantum Hall conductivity of the system by the so-called TKNN formula Eq. \eqref{conductivity}, \cite{TKNN}. Moreover, the presence of conducting edge states in topological insulators, or Majorana fermions in a topological superconductor, can be also related to the fact of having a topological non-trivial structure in the system.

We shall consider a time-reversal broken two-band system with the Fermi energy between both bands. The Chern number is given by
\begin{equation}\label{ChernN}
{\rm Ch}:=\frac{1}{2\pi}\int_{\rm BZ}d^2\bm{k} F_{xy}(\bm{k}),\quad F_{xy}(\bm{k}):=\partial_{x}A_{y}^{\rm B}(\bm{k})-\partial_{y}A_{x}^{\rm B}(\bm{k}),
\end{equation}
where BZ stands for Brillouin zone, $A_j^{\rm B}(\bm{k})={\rm i} \langle u_{\bm{k}}|\partial_{j} u_{\bm{k}}\rangle$ is the Berry connection and $|u_{\bm{k}}\rangle$ is the eigenvector corresponding to the lower energy band. This number is a topological invariant which only takes on integer values. While this kind of constructions can be extended to higher dimensional systems or systems with time-reversal symmetry \cite{LibroBernevig}, when attempting the generalization to density matrices via the Uhlmann connection, one finds the following  fundamental obstruction.

A natural way to generalize the Chern number to arbitrary density matrices would be to consider the first Chern class associated to the Uhlmann curvature, which is constructed from the Uhlmann connection $A$ via the standard formula for the non-Abelian case, Eq.\eqref{FnonAbelian},
\begin{equation}
F_{xy}^{\rm U}=\partial_{x}A_{y}-\partial_{y}A_{x}+[A_{x},A_{y}].
\label{FnA}
\end{equation}

Then, using the theory of characteristics classes \cite{Eguchi,Nakahara}, the (first) Chern number of the Uhlmann curvature would be given by
\begin{equation}
{\rm Ch}_{\rm U}:=\frac{{\rm i}}{2\pi}\int_{\rm BZ}d^2\bm{k} \Tr (F^{\rm U}_{xy}).
\label{ChnA}
\end{equation}
There is however, a problem with this construction, as ${\rm Ch}_{\rm U}$ turns out to be always zero. The reason for this is twofold:
\begin{enumerate}
\item First of all, the Uhlmann connection belongs to the $\mathfrak{su}(n)$ Lie algebra, so its trace vanishes and so does the trace of its curvature. Hence, using Eq. \eqref{ChnA}, ${\rm Ch}_{\rm U}$ is zero independently of the associated state $\rho_k$. This has more to do with the mathematical structure of the Uhlmann construction, than with physical properties of $\rho_k$.
\item On the other hand, the Chern number is 0 as long as there is a smooth gauge defined along the whole BZ \cite{Eguchi,Nakahara}. Note that this is the case for the Uhlmann $^{\rm U}(n)$ gauge, as we can always take the gauge $w_{\bm k}=\sqrt{\rho_{\bm{k}}}$ which is well defined provided that
$\rho_{\bm{k}}$ is not singular at some crystalline momentum $\bm{k}$. This is a rather natural condition and the usual case in physical situations.
\end{enumerate}

Additionally, the second condition, also implies that higher order Chern numbers vanish for the Uhlmann connection. Thus, ${\rm Ch}_{\rm U}=0$ in any case. This result was first proven in Ref. \cite{Viyuela_et_al14_2D}. 

In conclusion, this makes not obvious the extension of 2D topological invariants by means of the Uhlmann approach. We will show specifically the way to avoid these obstacles and define a proper topological invariant for general density matrices associated to these topological systems.

\begin{figure}[t]
\centering
\includegraphics[width=0.6\textwidth]{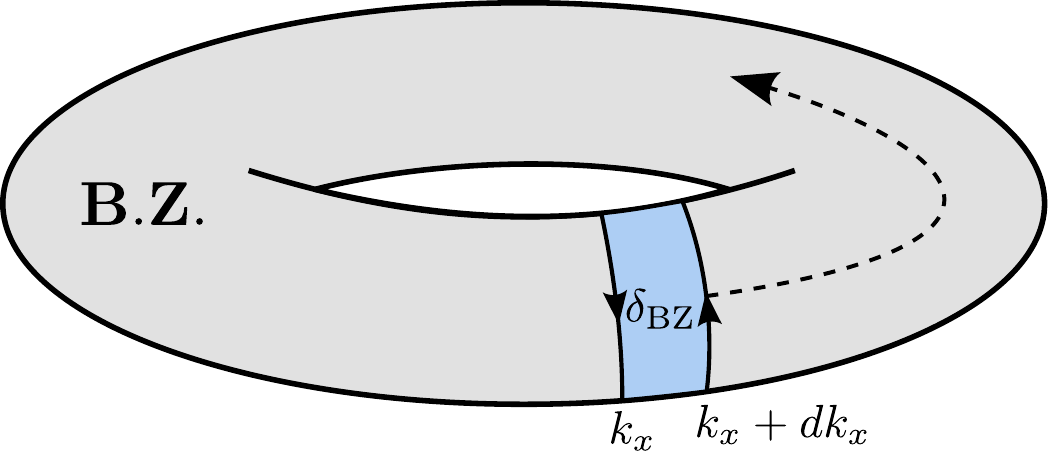}
\caption{This 2D torus represents the Brilloiun Zone (BZ) where the crystalline momentum of the system leaves. The figure represents the partition of the torus in infinitesimal slices of width $dk_x$, computing the Berry (or Uhlmann) phase associated to each slice and integrating of the whole homological cycle in $k_x$, one obtains the Chern number (Eq. \eqref{ChernW}) or the Uhlmann number (Eq. \eqref{nU}) respectively.}
\label{toro}
\end{figure}

\subsection{Topological Uhlmann numbers}

In this section, we explain how to construct single topological invariants associated to density matrices of 2D topological insulators and superconductors. The key tool for that is going to be the topological Uhlmann phase previously derived. The 2D construction as explained here was originally proposed in Ref. \cite{Viyuela_et_al14_2D} and also equivalently in Ref. \cite{Arovas14}.
A more detailed derivation and discussion not present in previous works is carried out in what follows.

In the previous subsection, we showed that the Chern number associated to the Uhlmann connection was always zero. However, the fact that ${\rm Ch}_{\rm U}$ becomes identically zero and the associated fiber bundle becomes trivial, does not imply that all topological properties of density matrices are trivial. If this assertion were true, we could not claim that systems at $T=0$ can be topologically ordered, as they are just a particular case of generally mixed density matrices. And in fact, we see that systems at $T=0$ display topological order in many fashions (non-zero Chern numbers, topological edge states, Majorana fermions, magneto-electric effect, ...). The key point in the discussion is that ${\rm Ch}_{\rm U}$ is not the only topological invariant that we can construct on a torus out of a density matrix.

With that in mind, let us analyst the problem at $T=0$ from a different perspective. Actually, in the Berry case, the Chern number \eqref{ChernN} can be rewritten as \cite{Abanin_et_al_13,Resta,Xiao}
\begin{equation} \label{ChernW}
{\rm Ch}=\frac{1}{2\pi}\oint dk_x \frac{d\Phi_{\rm B}(k_x)}{d k_x},
\end{equation}
where $\Phi_{\rm B}(k_x)=\oint dk_y A_y^{\rm B}(k_x,k_y)$ is the Berry phase along the $k_y$-nontrivial homological circle of the torus at the point $k_x$, and $\oint dk_x$ denotes the integration along the $k_x$-nontrivial homological circle. See Fig. \ref{toro}.

To prove the equality \eqref{ChernW} from the general definition of the Chern number, one divides the surface integral in Eq. \eqref{ChernN} into small slices along the $k_x$ direction,
\begin{equation}\label{ChernN2}
\frac{1}{2\pi}\int_{\delta_{\rm  BZ}}d^2\bm{k} F_{xy}(\bm{k}),
\end{equation}
where $\delta_{\rm  BZ}$ is an infinitesimal element of area that  is limited by one $k_y$-nontrivial homological circle of the torus at the point $k_x$ and another one at the point $k_x+dk_x$, as shown in Fig. \ref{toro}. The U(1)-gauge, which may be ill-defined over the whole BZ, is always well-defined in a sufficiently small slice, so that the Stokes' theorem is applicable. Thus,
\begin{equation}\label{ChernN3}
\frac{1}{2\pi}\int_{\delta_{\rm  BZ}}d^2\bm{k} F_{xy}(\bm{k})=\oint_{\partial\delta_{\rm  BZ}}A^{\rm B}(\bm{k})d\bm{k}=-\Phi_{\rm B}(k_x)+\Phi_{\rm B}(k_x+dk_x)=\frac{\partial\Phi_{\rm B}(k_x)}{\partial k_x}dk_x.
\end{equation}
Then, in the limit of slices with infinitesimal width, summing the slices at both sides of Eq. \eqref{ChernN3} is equivalent to performing the integral and one easily gets Eq. \eqref{ChernW}.

Note that, Eq. \eqref{ChernW} gives indeed a lot of information. If $\Phi_{\rm B}(k_x)$ is a continuous function and $\Phi_{\rm B}(k_x\rightarrow0)=\Phi_{\rm B}(k_x\rightarrow2\pi)$, then we trivially get ${\rm Ch}=0$. If $\Phi_{\rm B}(k_x)$ displays some $2\pi$-discontinuous jumps along the $k_x$ circle, then we can always write
\begin{equation}
\Phi_{\rm B}(k_x)={\rm e}^{{\rm i}2\pi\Theta{(k_x-k^c_x)}}\hat{\Phi}_{\rm B}(k_x),
\end{equation}
where $\hat{\Phi}_{\rm B}(k_x)$ is a smooth function equal to $\oint dk_y A_y^{\rm B}(k_x,k_y)$ mod. $2\pi$, $k^c_x$ is the point at which the discontinuity happens and $\Theta$ denotes the Heaviside step function. This way we can compute the derivative and perform the integration along the $k_x-$nontrivial homological circle of the torus. Actually, the Chern number is measuring the number of these $2\pi$ jumps, which is equivalent to the number of windings of $\Phi_{\rm B}(k_x)$ as the $k_x$ circle is covered. This is indeed a topological invariant. In particular a winding number, that associates every state of the system with an homotopy class of the Berry phase mapping
\begin{equation}
\Phi_{\rm B}(k_x): S^1\rightarrow S^1,
\end{equation}
between the nontrivial homological circle $S^1$ along $k_x$ and the complex Berry phases $U(1)\cong S^1$.

\begin{figure}[t]
\centering
\includegraphics[width=0.8\textwidth]{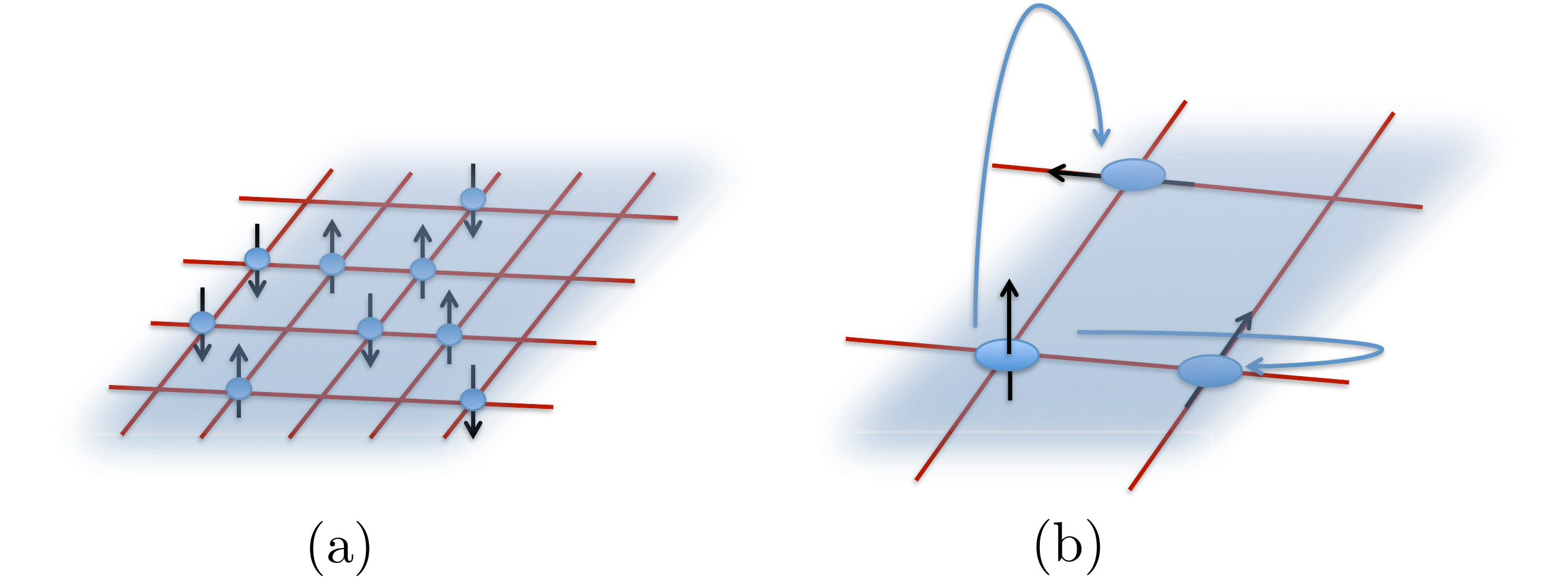}
\caption{A pictorial image of the QWZ model in real-space is shown here. In (a), we represent the on-site $\sigma_z$ part of the Hamiltonian interaction, where the arrows represent the spins in $z-$direction. In (b), we show an example of the effect of the spin-dependent hopping in the interaction. In particular, $\ket{\uparrow}_z~\longrightarrow \ket{\uparrow}_y$ when hopping along the $x-$axis, but $\ket{\uparrow}_z~\longrightarrow \ket{\downarrow}_x$ when hopping along the $y-$axis.}
\label{QWZ_lattice}
\end{figure}

As we have seen before, Eq. \eqref{ChernN} does not admit a direct generalization to density matrices, however, the equivalent formula \eqref{ChernW} for the Chern number at $T=0$ and pure states allows for a nontrivial generalization to density matrices.
To that aim, we will just replace the Berry phase $\Phi_{\rm B}(k_x)$ by the Uhlmann phase $\Phi_{\rm U}(k_x)$, Eq. \eqref{phiU}, in Eq. \eqref{ChernW}:
\begin{equation}\label{nU}
{\rm n}_{\rm U}:=\frac{1}{2\pi}\oint dk_x \frac{d\Phi_{\rm U}(k_x)}{d k_x}.
\end{equation}
Note that, many of the properties of the Berry case are kept. This number ${\rm n}_{\rm U}$ is also an integer by the same arguments given previously for the Chern number. If the $\Phi_{\rm U}(k_x)$ is a continuous function along the $k_x$ circle and $\Phi_{\rm U}(k_x\rightarrow0)=\Phi_{\rm U}(k_x\rightarrow2\pi)$, then we trivially obtain ${\rm n}_{\rm U}=0$. Otherwise, ${\rm n}_{\rm U}$ measures again the number of $2\pi$-discontinuous jumps along the $k_x$-nontrivial homological circle.
It is also trivially gauge invariant, as the Uhlmann phase is gauge invariant already.

This integer number is therefore a topological invariant which classifies the density matrices of a quantum system according to the homotopy class of the Uhlmann phase mapping,
\begin{equation}
\Phi_{\rm U}(k_x): S^1\rightarrow S^1,
\end{equation}
between the nontrivial homological circle $S^1$ along $k_x$ and the complex Uhlmann phases $U(1)\cong S^1$.

Furthermore, the limit at zero temperature is also well-defined. For pure states, the Uhlmann phase coincides with the Berry phase $\Phi_{\rm U}=\Phi_{\rm B}$. Thus, if we compute ${\rm n}_{\rm U}$ in a thermal (Gibbs) state, then the following relation is trivially satisfied
\begin{equation}
{\rm n}_{\rm U} \xrightarrow{T\rightarrow0} {\rm Ch},
\end{equation} hence, the generalization is faithful.

Additionally, since $\Phi_{\rm U}$ is an observable and we even discussed certain general schemes to measure it in different platforms \cite{Viyuela_et_al14}, then ${\rm n}_{\rm U}$ is also an observable. We will refer to this topological invariant ${\rm n}_{\rm U}$ as the \emph{Uhlmann number}.

As was shown in Ref. \cite{Viyuela_et_al14_2D}, the Uhlmann number characterizes representative models for topological insulators and superconductors in 2D. Remarkably, in models with high Chern number, one finds novel thermal-topological transitions between two nontrivial phases solely driven by temperature.

In what follows, we will analyze a new model for a two-band topological insulators at finite temperature according to the topological invariant ${\rm n}_{\rm U}$.

\subsection{QWZ Chern insulator.}

As an example of the applicability of this method, we present a model for a 2D topological insulator proposed in \cite{QWZ_06,QHZ_08}. Although the model is mathematically simple, it contains the main futures of the physics of a Chern insulator.

\begin{figure}[t]
\centering
\includegraphics[width=0.5\textwidth]{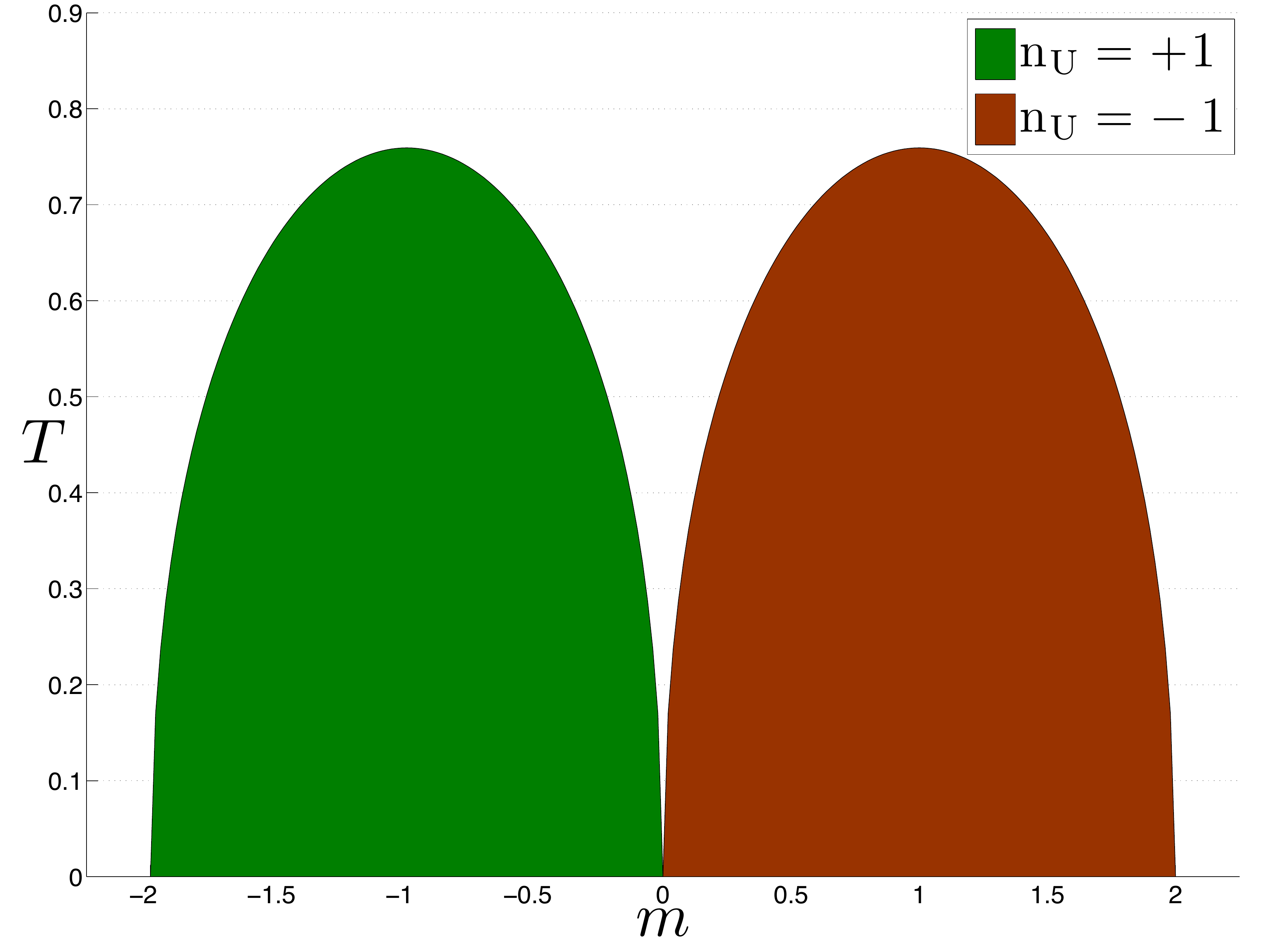}
\caption{Topological phase diagram given by the Uhlmann number for the QWZ model. Varying both the energy imbalance $m$ and the temperature $T$ we can go from a topological region to a trivial one. Note that ${\rm n}_{\rm U}$ remains quantized everywhere. If we fix $m$ and vary the temperature $T$, there is always a critical point where ${\rm n}_{\rm U}$ goes discontinuously and abruptly to zero.}
\label{Fig_QHZ}
\end{figure}

The model consist of a simple square lattice of spin-1/2 fermions, with spin-dependent nearest-neighbour hoppings, and an energy imbalance between spin up and down (see Fig. \ref{QWZ_lattice}). The Hamiltonian in real space is given by
\begin{equation}
H=\sum_{\vec{r}}\Big[c^{\dagger}_{\vec{r}}\frac{\sigma_z - {\rm i}\sigma_x}{2}c_{\vec{r}+\vec{x}}+c^{\dagger}_{\vec{r}}\frac{\sigma_z - {\rm i}\sigma_y}{2}c_{\vec{r}+\vec{y}} +\text{h.c.}\Big]+m\sum_{\vec{r}}c^{\dagger}_{\vec{r}}\sigma_zc_{\vec{r}}.
\label{Hs_r}
\end{equation}
where $c^{\dagger}_{\vec{r}},c_{\vec{r}}$ are fermionic creation and annihilation operators and $\vec{x}$ and $\vec{y}$ are unitary vectors along the $x$ and $y$ direction respectively. The Pauli matrices $\sigma_{x,y,z}$ act on the spin-1/2 degree of freedom.
The model describes the quantum anomalous Hall effect. The $\sigma_x$ and $\sigma_y$ terms appear due to strong spin-orbit coupling and the $\sigma_z$ part represents some ferromagnetic polarization. It has been shown to be realisable in ${\rm Hg}_{1-x}{\rm Mn}_x{\rm Te}/{\rm Cd}_{1-x}{\rm Mn}_x{\rm Te}$ quantum wells with a proper amount of Mn spin polarization \cite{Liu_08}.

In Fig. \ref{QWZ_lattice} we sketch the dynamics of the system for a certain spin orientation as an example. Figure \ref{QWZ_lattice}(a) represents the on-site $\sigma_z$ part, and Fig. \ref{QWZ_lattice}(b) shows the hopping mechanism for a certain spin state in order to visualise the dynamics. When a fermion with spin up in the $z-$direction hops along the $x-$direction, then $\ket{\uparrow}_z~\longrightarrow \ket{\uparrow}_y$. If it hops along the $y$-direction, then $\ket{\uparrow}_z~\longrightarrow \ket{\downarrow}_x$, as it is shown in Fig. \ref{QWZ_lattice}(b).

Taking periodic boundary conditions the Hamiltonian in momentum space reads
\begin{equation}
H({\bm k})=\sin{(k_x)}\sigma_x+\sin{(k_y)}\sigma_y +[m+\cos{(k_x)}+\cos{(k_y)}]\sigma_z,
\label{H_QWZ}
\end{equation}
where $m$ is some energy imbalance between to atoms of different species.

At $T=0$, the different topological phases as classified by the Chern number, Eq. \eqref{ChernN}, are:
\begin{equation}
{\rm Ch}= \begin{cases}
+1, & \text{if}~-2<m<0, \\
-1, & \text{if}~0<m<2, \\
0, & \text{if}~\text{otherwise}. \\
\end{cases}
\label{C_QWZ}
\end{equation}

For nontrivial regions ${\rm Ch}=\pm1$, the system presents metallic edge states at the boundary. Indeed, the conductivity of these edge states is given by the Chern number $\sigma_{{\rm Edge}}=\frac{e^2}{h}{\rm Ch}$.

At finite temperature, the different topological phases as classified by the Uhlmann number, Eq. \eqref{nU}, are graphically represented Fig. \ref{Fig_QHZ}. The system displays nontrivial topological phases ${\rm n}_{\rm U}=\pm1$ even at nonzero temperature provided it is below a certain critical value $T_c$, where ${\rm n}_{\rm U}$ goes to zero. This critical temperature $T_c$ reaches its maximum at the points $m=\pm1$ within the topological phases ${\rm Ch}=\pm1$ at $T=0$. These points are associated with the highest value of the gap. This matches with the common intuition that the resilience of topological order has to do with the ratio between the gap and the temperature, and the larger the gap, the more stable it is. As expected, in the limit of $T=0$ we recover the same topological diagram as given by the Chern number.

\begin{figure}[t]
\centering
\includegraphics[width=0.8\textwidth]{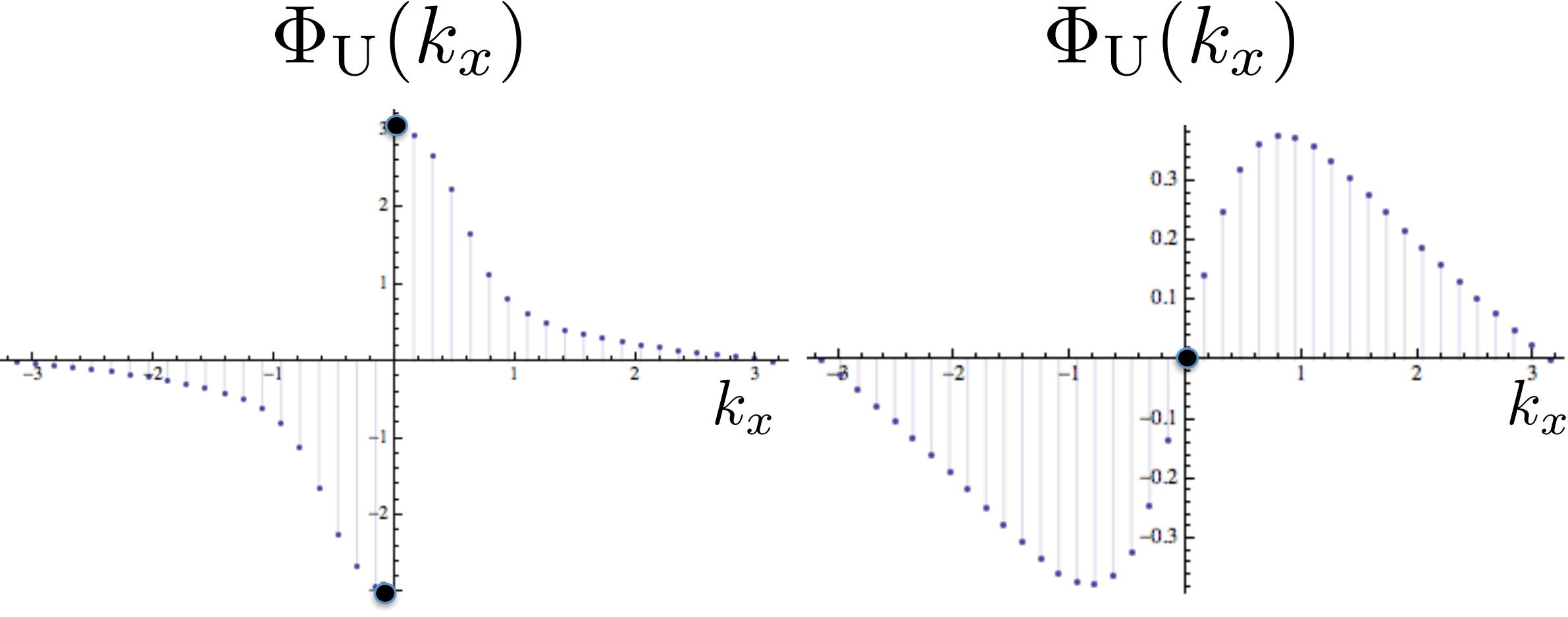}
\caption{We represent $\Phi_{\rm U}(k_x)$ in the QWZ model for $m=-1$. On the left figure, we set $T=0.1$ and observe a discontinuous jump in the phase which is responsible for the robustness of the topological phase at finite temperature. On the right figure, we take a higher temperature $T=1.2$ where the system is not topological anymore, and the discontinuity in the phase disappears.}
\label{FigPhiU}
\end{figure}

Actually, we can understand even better the discontinuity in the Uhlmann number when crossing different topological regions looking at Fig. \ref{FigPhiU}. We represent $\Phi_{\rm U}(k_x)$ for different topological sectors. Namely, we have set $m=-1$ which is a topological region at $T=0$. At low temperature we see that there is still a jump in the phase at a certain critical momentum $k_c=0$, which accounts for ${\rm n}_{\rm U}=1$. However, in the same figure we can see what happens at higher temperature where the jump disappears, the Uhlmann phase is a continuous function of $k_x$ and hence, ${\rm n}_{\rm U}=0$. This supports the previous explanation that the topological Uhlmann number accounts for the number of discontinuous jumps of the Uhlmann phase in momentum space.

Thus, we see that thermal topological phase transitions are not a unique phenomenon of the 1D case \cite{Viyuela_et_al14} and they are also found in two-dimensional systems.

\section{CONCLUSIONS}
\label{sec:V}

In this paper we have characterized a new emblematic model for a topological insulator, the Qi-Wu-Zhang model introduced in \cite{QWZ_06,QHZ_08} using the topological Uhlmann number introduced in \cite{Viyuela_et_al14_2D} in order to find a
stable SPT phase under external thermal fluctuations in two-dimensions. This was first originally discovered
in emblematic examples of topological insulators and superconductors in one-dimension \cite{Viyuela_et_al14_2D}, which later
lead to higher dimensional constructions \cite{Arovas14,Viyuela_et_al14_2D}. A complete phase diagram of the topological Uhlmann number as a function of the parameters of the Hamiltonian in the  Qi-Wu-Zhang  and as a function of the temperature of the system has been shown in Fig. \ref{Fig_QHZ}. We clearly see stable thermal topological phases in large regions of the phase diagram
where the topological invariant remains quantized and critical temperatures where the system jumps onto a trivial phase with zero
value for the topological Uhlmann number.

In order to apply our newly developed topological invariant \cite{Viyuela_et_al14_2D} to the topological insulator in 2D given by
the Qi-Wu-Zhang model, we have presented a self-contained explanation of the theory of geometric phases in quantum mixed
states represented by density matrices and how they may account for new topological effects under certain conditions.
Thus, we have introduced the main concepts and results of the Uhlmann geometric theory. We have used the Uhlmann geometric phase as a tool to characterise SPT systems like topological insulators and superconductors at finite temperature. The Uhlmann phase acquires a topological character in this situations remaining quantized even for a finite range of temperature. In fact, when computed  for a thermal state of  generic two-band models in 1D, the complicated expression of the Uhlmann phase simplifies significantly  and can be related to the winding number of the Hamiltonian and the temperature of the system. It has also a well-defined limit at $T=0$, tending to the Berry/Zak phase, which makes the theory consistent and faithful.

Morover, we have presented how this tool can be generalised to 2D systems in a detailed fashion. The topological invariant in this case is the topological Uhlmann number, which generalises the Chern number for symmetry-protected time-reversal-broken topological insulators and superconductors in 2D. In fact, with these results, we may think of the possibility of extending the current classification of topological insulators and superconductors on several spatial dimensions \cite{Ludwig,Kitaev_2009} (also called the ``periodic table"), to the situation of thermal SPT states with fermions using topological Uhlmann numbers.

Let us emphasise  that the existence of critical temperatures seems rather natural in the Uhlmann approach.
We find that for thermal states, this approach sets on equal footing the temperature and the coupling constants in the original Hamiltonian. Therefore, if there is a critical behavior as a function of tunnelings and/or staggered potentials, then one should obtain a critical behavior with temperature as well. Moreover, since by raising the temperature $T$, the quantum coherence properties of any state are affected, it is rather natural that the Uhlmann number decreases by warming the system up.

Beyond the scope of this paper it is the possible adaptation of the Uhlmann formalism to explore interacting topological phases at finite temperature. They include not only bosonic/spin topological phases but also interacting fermionic SPT phases that have been recently proposed and classified \cite{Gu_09,Chen_12,Gu_14} based on group cohomology and supercohomology theory, and response field theory corresponding to a non-trivial fiber bundle of the symmetry group \cite{Cheng_14,Wang_15}. In this regard, the application of topological Uhlmann invariants to more complicated models and higher dimensions may lead to novel thermal topological transitions and new effects.

\begin{acknowledgments}

 We thank the Spanish MINECO grant FIS2012-33152, FIS2009-10061, CAM research consortium QUITEMAD$+$ S2013/ICE-2801, European Commission
PICC: FP7 2007-2013 grant No.~249958, UCM-BS grant GICC-910758, FPU MEC grant and Residencia de Estudiantes. 

\end{acknowledgments}

\end{document}